\documentclass[letterpaper,twocolumn,10pt]{article}
\usepackage{usenix,epsfig,endnotes}
\usepackage{amsmath, amsfonts, bbm, amsthm}
\usepackage{makecell}
\usepackage[table]{xcolor}
\usepackage{threeparttable}
\usepackage{graphicx}
\usepackage{subfig}
\usepackage{mathrsfs}
\usepackage{bm}

\begin{document}

\newtheorem{theorem}{Theorem}[section]
\newtheorem{lemma}{Lemma}[section]
\newtheorem{definition}{Definiton}[section]
\newtheorem{corollary}{Corollary}[section]
\newtheorem{claim}{Claim}[section]
\newtheorem{proposition}{Proposition}
\newtheorem{conjecture}{Conjecture}[section]
\newtheorem{property}{Property}[section]
\newtheorem{example}{Example}
\newtheorem{assumption}{Assumption}

\date{}

\title{\Large \bf Strict Optimality of Frequency and Distribution Estimation Under Local Differential Privacy}

\author{
{\rm Mingen Pan}\\
Google \\
mepan94@gmail.com
}

\maketitle

\thispagestyle{empty}

\subsection*{Abstract}
This paper establishes the strict optimality in precision for frequency and distribution estimation under local differential privacy (LDP). We prove that a linear estimator with a symmetric and extremal configuration, and a constant support size equal to an optimized value, is sufficient to achieve the theoretical lower bound of the $\mathcal{L}_2$ loss for both frequency and distribution estimation. The theoretical $\mathcal{L}_1$ lower bound is also achieved asymptotically. Furthermore, we derive that the communication cost of such an optimal estimator can be as low as $\log_2(\frac{d(d-1)}{2}+1)$ bits, where $d$ denotes the dictionary size, and propose an algorithm to generate this optimal estimator.

In addition, we introduce a modified Count-Mean Sketch and demonstrate that it is practically indistinguishable from theoretical optimality with a sufficiently large dictionary size (e.g., $d=100$ for a privacy parameter of $\epsilon = 1$). We compare existing methods with our proposed optimal estimator to provide selection guidelines for practical deployment. Finally, the performance of these estimators is evaluated experimentally, showing that the empirical results are consistent with our theoretical derivations.

\section{Introduction}
The frequency of a dataset and the distribution of a population represent the most fundamental statistics. They serve as the essential building blocks for deriving a wide array of statistical properties, ranging from basic mean and variance to more complex higher-order moments.

While computing frequencies or distributions is straightforward when an analyst has unrestricted access to a dataset, typically involving a simple iteration using a hash map, the task becomes significantly more complex when raw individual values are inaccessible. This is often the case when robust data privacy protections are implemented to mask sensitive information.

This paper focuses on a privacy-preserving technique known as Local Differential Privacy (LDP) \cite{LDP}, the localized counterpart to Differential Privacy (DP) \cite{dwork2006calibrating}, which has become the \textit{de facto} standard for data privacy. Before reporting an aggregate statistic from a dataset, DP injects sufficient noise into the result to ensure that an adversary cannot determine whether a specific data point was included in the original set. However, DP traditionally requires centralized data processing, meaning it collects and stores raw data from clients before adding noise to the final aggregation. This architecture leaves data vulnerable to server breaches and eavesdropping during the transmission of raw values.

LDP was proposed to mitigate these risks by perturbing data locally on the client side before it is sent to the server. Consequently, even if intercepted, an adversary cannot effectively infer the original value from the perturbed output. The degree of this perturbation is quantified by a privacy parameter, $\epsilon$, representing the maximum amount of knowledge an adversary can gain from the perturbed result \cite{pan2023knowledge}.

Since frequency and distribution are such fundamental statistical properties, numerous LDP algorithms have been developed to estimate them effectively. These algorithms are discussed in detail in Section \ref{sec:previous_work}. Among them, the Subset Selection mechanism proposed by \cite{wang2016mutual, ye2018subset} achieved state-of-the-art precision regarding $\mathcal{L}_1$ and $\mathcal{L}_2$	losses. These metrics, formally defined in Section \ref{sec:frequency}, measure the overall precision across all estimated frequencies or distributions. Notably, since the introduction of Subset Selection, no subsequent LDP estimator has surpassed its leading precision in these metrics. 

While the leading precision has remained unsurpassed for years, it remains unknown whether these algorithms are strictly optimal. The Subset Selection paper \cite{ye2018subset} established a lower bound for LDP distribution 
estimation, proving that their algorithm is order-optimal when $e^{\epsilon} \ll d$. However, a significant gap remains in the constant term; specifically, the derived lower bound can be as low as $\frac{1}{512}$ of the achieved precision. 
A follow-up study \cite{ye2019optimal} suggested that Subset Selection achieves asymptotic strict optimality for distribution estimation, but this does not apply to finite datasets or frequency estimation (see Section \ref{sec:previous_work}). It remains unclear whether an algorithm exists to bridge this gap, or if current approaches are already strictly optimal.

This paper directly addresses these questions by establishing a strict lower bound for LDP frequency and distribution estimation precision in terms of $\mathcal{L}_1$ and $\mathcal{L}_2$ losses:
\begin{proposition}
Let $d$, $\epsilon$, $n$, and $\hat{f}$ denote the dictionary size (the number of all possible inputs), privacy parameter, dataset size, and frequency estimator, respectively. The minimum $\mathcal{L}_2$ loss is given by
\begin{equation*}
\min_{\hat{f}} \mathcal{L}_2(\hat{f}) =
\begin{cases}
\displaystyle \frac{(d - 1) [4 d e^{\epsilon} - (e^{\epsilon} + 1)^2]}{n d (e^{\epsilon} - 1)^2} & \text{if } d \ge e^{\epsilon} + 1,  \\[15pt]
\displaystyle \frac{ (d - 1) (d + 2 e^{\epsilon} - 2) }{ n (e^{\epsilon} - 1)^2 } & \text{otherwise.}
\end{cases}
\end{equation*}
Meanwhile, $\min_{\hat{f}}\mathcal{L}_1(\hat{f}) \approx \sqrt{ \frac{2}{\pi}\min_{\hat{f}}\mathcal{L}_2(\hat{f})}.$
\end{proposition}

\noindent Full details are presented in Theorem \ref{thm:optimized_l1_l2} and Corollary \ref{cor:l1_l2_k=1}. Distribution estimation has a similar result: $\min_{\hat{\theta}} \mathcal{L}_2(\hat{\theta}, n) = \min_{\hat{f}} \mathcal{L}_2(\hat{f}, n) + \frac{1 - 1/d}{n}$, and $\min_{\hat{\theta}}\mathcal{L}_1(\hat{\theta}, n) \approx \sqrt{ \frac{2}{\pi}\min_{\hat{\theta}}\mathcal{L}_2(\hat{\theta}, n)}$, where $\hat{\theta}$ denotes a distribution estimator.

To establish these tight lower bounds, this paper first demonstrates that an estimator constructed from a uniformly random permutation (URP), when paired with an optimized reconstruction matrix, achieves the theoretical $\mathcal{L}_2$ loss lower bound derived from the Fisher information for any given perturbation matrix. The theoretical $\mathcal{L}_1$ lower bound is also achieved asymptotically. We then prove that these losses are further minimized when the perturbation matrix maintains a constant support size across all responses. Under these conditions, we express $\mathcal{L}_1$ and $\mathcal{L}_2$ as functions of the support size while holding other parameters constant. Finally, we determine the optimal support size to derive the strict lower bounds for both loss functions. The complete proof is provided in Section \ref{sec:strict_optimality}.

Furthermore, we derive that the communication cost, defined as the number of bits required to transfer a response from a client, for an optimal estimator is upper bounded by $\log_2(\frac{d(d-1)}{2} + 1)$ bits, where $d$ denotes the dictionary size. We then develop an optimal estimator, Weighted Subset Selection, which achieves this reduced communication cost despite having high precomputation requirements. Subsequently, we evaluate the optimality of existing algorithms and find that Subset Selection \cite{ye2018subset} already achieves strict optimality in terms of precision, while its communication cost is linear in $d$. We also demonstrate that the Count-Mean Sketch (CMS), with minor modifications, can approach strict optimality provided the dictionary size is sufficiently large, such as when $d = 100$ for $\epsilon = 1$. Notably, this modified version, which we refer to as Optimized CMS, requires a significantly lower communication cost of $O(\log_2(d))$ bits, making it much more efficient than Subset Selection. Consequently, we provide practical guidance for their deployment; the rule of thumb is to employ Optimized CMS when the dictionary size is sufficiently large and to use either Weighted Subset Selection or the original Subset Selection otherwise.

Finally, we conduct two experiments in Section \ref{sec:experiment} to compare the performance of the optimal algorithms discussed in Section \ref{sec:optimized_algorithms} with the theoretical strict lower bounds of $\mathcal{L}_1$ and $\mathcal{L}_2$. Both experiments demonstrate that all three optimal algorithms align perfectly with the strict lower bounds.

\subsection{Previous Work}
\label{sec:previous_work}
Numerous LDP algorithms have been developed for frequency and distribution estimation. Randomized Response \cite{warner1965randomized, kairouz2014extremal}, predating the formal LDP definition, was initially designed for the frequency estimation of sensitive data. RAPPOR \cite{rappor} later achieved estimation precision for individual values independent of the dictionary size $d$, although its communication cost remains linear in $d$. Subsequently, methods such as Hadamard Encoding \cite{bassily2015hadamard}, private Count-Mean Sketch \cite{apple_privacy}, Local Hashing and asymmetric RAPPOR\cite{wang2017localhashing}, Subset Selection \cite{ye2018subset}, Hadamard Response families \cite{acharya2019hr, chen2020rhr}, Projective Geometry \cite{feldman2022private}, Block Design \cite{park2024exactly}, Optimized Count-Mean Sketch \cite{pan2025improving}, and Modular Subset Selection \cite{arcolezi2026private} were developed to enhance precision, with several reducing the communication cost to $O(\log d)$. Among these, Subset Selection \cite{ye2018subset} achieves state-of-the-art precision for $\mathcal{L}_1$ and $\mathcal{L}_2$ metrics, while some algorithms \cite{feldman2022private, pan2025improving} asymptotically approach this performance.


Several existing studies analyze the optimality of LDP distribution estimation. Kairouz et al. (2014) \cite{kairouz2014extremal} demonstrated that any optimal LDP algorithm can be converted into an extremal configuration. While Ye and Barg (2018) \cite{ye2018subset} established lower bounds for $\mathcal{L}_1$ and $\mathcal{L}_2$ losses when $e^{\epsilon} \ll d$, these bounds are not tight and have constant terms that are significantly smaller than the strict bounds derived in this work. A follow-up study \cite{ye2019optimal} proved that Subset Selection achieves asymptotic strict optimality for $\mathcal{L}_1$ and $\mathcal{L}_2$ in distribution estimation. However, their proof relies on numerous approximations requiring an extremely large dataset size $n$. Furthermore, they do not address how Subset Selection deviates from optimal bounds when $n$ is finite. Finally, their proof of optimality for distribution estimation, which involves the probability mass function (PMF) of a multinomial distribution, does not directly translate to frequency estimation; the latter involves estimating a Poisson multinomial distribution, the PMF of which lacks a concise closed-form expression. In contrast, our paper derives the strict $\mathcal{L}_2$ lower bound for any dataset size in both frequency and distribution estimation, and provides a sufficient condition to achieve this bound.

\section{Background}

\subsection{Local Differential Privacy (LDP)}
\label{sec:ldp}
Local Differential Privacy (LDP) was initially proposed in \cite{LDP} to ensure that querying an individual object does not significantly disclose its actual value. Consider an object $X \in [d] = \{1, 2, .., d\}$ being queried by a mechanism $M$:

\begin{definition}
   $M$ is $\epsilon$-LDP if and only if $\forall x, x' \in [d]$, and $\forall o \in [m]$,
\begin{equation}
    \Pr(M(X) = o | X = x) \le e^{\epsilon} Pr(M(X) = o | X = x') , \notag
\end{equation} 
\end{definition}

\noindent where $m$ represents the output domain size of $M$. In this paper, we may abbreviate $(M(X) | X = x)$ as $M(x)$ when the specific object is not the primary focus and its value is denoted by $x$.

\subsection{Frequency Estimation}
\label{sec:frequency}

Denote $X^{(i)}$ to be the $i$-th object of a dataset of size $n$, where each object takes a value from a dictionary $[d]$. Thus, $\forall i \in [n]: X^{(i)} \in [d]$. The true frequency of a value $x$ is defined as $f(x) = \frac{\sum_{i \in [n]} \mathbbm{1}(X^{(i)} = x)}{n}$, where $\mathbbm{1}$ denotes an indicator function. Let $\mathbf{f}$ be the true frequency vector such that its $x$-th element represents $f(x)$.

Under LDP, the privacy of the original dataset is preserved by preventing direct access to its raw values. Instead, each client perturbs their original value using an LDP mechanism $M$ and reports the result to the server for frequency estimation. The process is defined as follows:

\begin{enumerate}
    \item Each client $X$ returns a response $O = M(X)$, represented as a one-hot encoding. If $O$ corresponds to a value $o$, its $o$-th element is set to one.
    \item The server collects all responses $[O^{(i)}]_{i \in [n]}$ and computes their frequency vector $\mathbf{f_O} = \frac{1}{n} \sum_{i=1}^{n} M(X^{(i)})$. The $o$-th element of this vector represents the frequency of the output $o$ across all responses.
    \item The estimated frequency vector is computed as $\mathbf{\hat{f}} = \hat{f}(\mathbf{f_O})$, where $\hat{f}$ is an estimator and the $x$-th element of $\mathbf{\hat{f}}$, denoted $\hat{f}(x)$, serves as an estimate of the true frequency $f(x)$.
\end{enumerate}

To evaluate the overall precision of an estimator $\hat{f}$, we define $L_1$ and $L_2$ norms as:
\begin{align*}
    L_1(\hat{f}, \mathbf{f}, n) &= \mathbb{E} \left[ \sum_{x=1}^d |\hat{f}(x) - f(x)| \right], \\
    L_2(\hat{f}, \mathbf{f}, n)  &=  \mathbb{E} \left[ \sum_{x=1}^d (\hat{f}(x) - f(x))^2 \right].
\end{align*}

Assuming the estimator is unbiased (i.e., $\forall x: \mathbb{E}[\hat{f}(x)] = f(x)$), we have $\mathbb{E}[(\hat{f}(x) - f(x))^2] = \text{Var}(\hat{f}(x))$. Consequently, the $L_2$ norm can be expressed as:

\begin{equation}
    L_2(\hat{f}, \mathbf{f}, n)  =  \sum_{x=1}^d \text{Var}(\hat{f}(x)). 
\end{equation}

\noindent Additionally, the worst-case norms (i.e., loss functions) are defined to be 
\begin{equation*}
    \mathcal{L}_{1|2}(\hat{f}, n) = \max_{\mathbf{f}}  L_{1|2}(\hat{f}, \mathbf{f}, n) , \label{eq:l2_def}
\end{equation*}
\noindent where $\mathcal{L}_{1|2}$ can be either $\mathcal{L}_{1}$ or $\mathcal{L}_{2}$. 


Let $P$ be the perturbation matrix of $M$, where $P_{o,x} = \Pr(M(x) = o)$. Since $\mathbb{E}[\mathbf{f_O}] = P \mathbf{f}$, a linear transformation $Q$ can reconstruct $\mathbf{f}$ from $\mathbf{f_O}$ if $QP = I$, where $I$ is the identity matrix.

\begin{definition}
     $\text{LN}(P, Q) = Q \mathbf{f_O} = \frac{1}{n} \sum_{i \in [n]} Q M(X^{(i)})$ denotes a linear estimator that uses $P$ to perturb and $Q$ to reconstruct.
\end{definition}

\noindent Given that $\mathbb{E}[Q \mathbf{f_O}] = Q P \mathbf{f} = \mathbf{f}$, we obtain the following result:

\begin{theorem}
    $\text{LN}(P, Q)$ is an unbiased frequency estimator.
\end{theorem}

\noindent Notably, $P$ must be of full column rank; otherwise, $\mathbf{f}$ cannot be reconstructed unbiasedly from $\mathbf{f_O}$. Since the size of the response space may exceed the dictionary size (i.e., $m \ge d$), there may exist multiple reconstruction matrices $Q$ such that $QP = I$.


We define $\text{Var}(\hat{f})$ as the variance vector of $\hat{f} = \text{LN}(P, Q)$, where $x$-th element represents $\text{Var}(\hat{f}(x))$. It can be formulated as follows:
\begin{theorem}
    \label{thm:var_for_any_P}
    \begin{equation*}
        \text{Var}(\hat{f}) = \frac{1}{n} [ (Q \circ Q) P  \mathbf{f} - \mathbf{f} ],
    \end{equation*}
    \noindent where $(Q \circ Q)$ denotes the point-wise (Hadamard) square, i.e., $(Q \circ Q)_{i, j} = Q_{i, j}^2$.
\end{theorem}

\noindent The proof is provided in Appendix \ref{sec:proof_thm:var_for_any_P}.

The relationship between $L_1$ and $\text{Var}(\hat{f}(x))$ is established in the following theorem:

\begin{theorem}
    \label{thm:l1_approx}
    The $L_1$ of $\hat{f} = \text{LN}(P, Q)$ is approximated as:
    \begin{equation}
        \lim_{n \rightarrow \infty} L_1(\hat{f}, \mathbf{f}, n) = \sum_{x \in [d]} \sqrt{\frac{2}{\pi} \text{Var}(\hat{f}(x))}.
    \end{equation}
\end{theorem}
\noindent The proof is provided in Appendix \ref{sec:proof_thm:l1_approx}.

\subsection{Distribution Estimation}
Assume that each object $X^{(i)}$ are independently sampled from an identical category distribution $\mathcal{C}(\bm{\theta})$, where the $x$-th element of $\bm{\theta}$ denotes the probability of generating value $x$. If the objects are perturbed by an LDP mechanism following Section \ref{sec:frequency}, the estimator $\text{LN}(P,Q)$ also estimates the distribution $\bm{\theta}$. Denote $\hat{\bm{\theta}}$ as the estimated distribution, which satisfies
\begin{theorem}
If $\hat{\theta} = \text{LN}(P, Q)$,
    \begin{align*}
        \mathbb{E}[\hat{\theta}] & = \bm{\theta} \\
        \text{Var}(\hat{\theta}) & = \frac{1}{n} [ (Q \circ Q) P  \bm{\theta} - \bm{\theta}^2  ]
    \end{align*}
\label{thm:var_theta}
\end{theorem}
\noindent The detailed proof is provided in Appendix \ref{sec:proof_thm:var_theta}.

\subsection{Extremal Configuration}
\label{sec:extremal}
An LDP perturbation mechanism is defined as \textit{extremal} if and only if each specific output has exactly two possible emission probabilities across all input values, and these probabilities maintain a ratio of $e^{\epsilon}$.

\begin{definition}
   $M$ is $\epsilon$-LDP and extremal if and only if $\forall x \in [d]$, and $\forall o \in \mathcal{O}$, the following condition holds:
\begin{equation}
    \Pr(M(x)= o) \in \{ p_o, e^{\epsilon} p_o \} , \notag
\end{equation} 

\noindent where $p_o$ is referred to as the base probability associated with response $o$.
\label{def:extremal}
\end{definition}

\begin{definition}
    For a response $o$ within an extremal configuration, the support set of $o$ is defined as the set of inputs $\{ x | \Pr(M(x)= o) = e^{\epsilon} p_o \}$. 
    
    The support size of $o$, denoted as $k_o$, represents the cardinality of this support set.
\label{def:support_set}
\end{definition}

In the subsequent sections, we define $S_{o, x}$ as an indicator function signifying whether $o$ supports $x$. Specifically, $S_{o, x} = 1$ if $x$ is an element of the support set of $o$, and $S_{o, x} = 0$ otherwise. This relationship can be expressed in matrix form as:
\begin{equation}
    P = P_O [(e^{\epsilon} - 1) S + J], \label{eq:extremal_matrix_form}
\end{equation}
\noindent where $P_O = \text{diag}(p_1, p_2, \dots, p_m)$ and $J$ is an $m \times d$ all-ones matrix.

The extremal configuration satisfies an invariant property that is foundational for the subsequent sections:

\begin{lemma}
If $M$ is an $\epsilon$-LDP extremal configuration, then
\begin{equation*}
    \forall x \in [d]: \sum_{o\in \mathcal{O}} S_{o, x} p_o = \frac{1 - \sum_{o \in \mathcal{O}} p_o}{e^{\epsilon} - 1} = \text{const}.
\end{equation*}
\label{lem:constant_sp}
\end{lemma}

This property is derived directly by summing the probabilities of all possible responses:

\begin{equation*}
       \forall x \in [d]: \sum_{o\in \mathcal{O}} S_{o, x} e^{\epsilon} p_o + (1 - S_{o, x}) p_o = 1 \qed
\end{equation*}

Notably, $p^* = e^{\epsilon} \sum_{o \in \mathcal{O}} S_{o, x} p_o$ is also referred to as the \textit{self-support probability} (Definition \ref{def:sym}). The extremal configuration ensures that the self-support probability remains constant across all original values.

Existing research has established the following:
\begin{theorem}
    Any estimator lacking an extremal configuration can be transformed into one that possesses such a configuration.
\end{theorem}
One may refer to \cite{kairouz2014extremal} for a detailed proof. An informal justification is that any non-extremal response can be decomposed into a collection of extremal responses. For instance, consider a scenario where the dictionary size $d = 3$. Suppose the probability of a response $o$ satisfies $[Pr(M(x)= o)]_{x \in [d]} = [1, k e^{\epsilon}, e^{\epsilon} ] \cdot p_o$, where $k \in (e^{-\epsilon}, 1)$. In this case, the response $o$ can be replaced by two new responses, $o_1$ and $o_2$, characterized by the following probability sets:
\begin{align*}
    & [\Pr(M(x)= o_1)]_{x \in [d]} = [1, e^{\epsilon}, e^{\epsilon} ] \cdot \frac{k e^{\epsilon} - 1}{e^{\epsilon} - 1} p_o   \\
    & [\Pr(M(x)= o_2)]_{x \in [d]}  = [1, 1, e^{\epsilon} ] \cdot \frac{e^{\epsilon} (1 - k)}{e^{\epsilon} - 1} p_o.
\end{align*}

An estimator can simply replace any occurrence of $o_1$ or $o_2$ as the response $o$, rendering it identical to the original reconstruction. Consequently, an optimal estimator with an extremal configuration always exists.

\subsection{Symmetric Configuration}
\label{sec:sym_config}
Wang et al. \cite{wang2017localhashing} reviewed the LDP frequency estimators available at that time and proposed a \textit{symmetric configuration} framework that encompasses all existing estimators. The formal definition is as follows:

\begin{definition}
   An $\epsilon$-LDP mechanism $M$ has a symmetric configuration, if and only if 
   \begin{align*}
       & \forall x \in [d]: \Pr(M(x)) \in \{ o | S_{o, x} = 1 \} = p^*. \\
       & \forall x, x' \in [d] \text{ and } x \ne x': \Pr(M(x')) \in \{ o | S_{o, x} = 1 \} = q^*. 
   \end{align*}
\noindent where $p^*$ and $q^*$ are constants independent of $x$ and $x'$, and $S$ is the support matrix of $M$. The values $p^*$ and $q^*$ are referred to as the self-support and cross-support probabilities, respectively.
\label{def:sym}
\end{definition}
\noindent This can be expressed equivalently in matrix form as:
\begin{equation}
    S^T P = (p^* - q^*) I + q^* J
\end{equation}
\noindent where $P$ is the perturbation matrix of $M$, $I$ is the identity matrix, and $J$ is the all-ones matrix. $S^T P$ is referred to as the \textit{support probability matrix}.

An estimator for symmetric configuration is proposed by \cite{wang2017localhashing} as:
\begin{equation}
    \hat{f}(x) = \hat{\theta}(x) = \frac{(\sum_{i \in [n]} S_{O^{(i)}, x}) / n - q^* }{p^* - q^*} ,
    \label{eq:sym_fx}
\end{equation}
\noindent  where $S_{O^{(i)}, x}$ is an indicator function denoting whether the response $O^{(i)}$ from the $i$-th object supports the value $x$. The support counts can also be represented in matrix form:
\begin{equation}
    [\sum_{i \in [n]}S_{O^{(i)}, x}]_{x \in [d]} =  S^T \sum_{i \in [n]} M(X^{(i)}) . \label{eq:S^T_M_matrix}
\end{equation}

\noindent The expectation and variance of this estimator are derived as follows:
\begin{theorem}
    \begin{align}
        & \mathbb{E}[\hat{f}] = \mathbf{f} \\
        & \text{Var}(\hat{f}) = \frac{q^* (1 - q^* )}{n (p^* - q^*)^2} + \mathbf{f} \frac{1 - p^* - q^*}{n (p^* - q^*) } . \label{eq:var_fx} \\
        & \text{Var}(\hat{\theta}) = \frac{q^* (1 - q^* )}{n (p^* - q^*)^2} + \bm{\theta} \frac{1 - p^* - q^*}{n (p^* - q^*) } + \frac{\bm{\theta} - \bm{\theta}^2}{n}. 
    \end{align}
    \label{thm:sym_exp_var}
\end{theorem}

$\mathbb{E}[\hat{f}]$ and $\text{Var}(\hat{f})$ were originally proven in \cite{wang2017localhashing}, while $\text{Var}(\hat{\theta})$ is derived similarly. For frequency estimation, the result can be verified by treating the term $\sum_{i \in [n]} S_{o^{(i)}, x}$ as the sum of two independent binomial random variables: $\text{Bin}(n f(x), p^*)$ and $\text{Bin}(n(1-f(x)), q^*)$. These variables represent the number of responses from the original value $x$ and supporting $x$, and the number of responses from the original value other than $x$ and supporting $x$, respectively. For distribution estimation, this term is instead modeled as $\text{Bin}(n, \theta(x) p^* + (1 - \theta(x)) q^*)$.

\subsection{Uniformly Random Permutation}
This section presents an approach to transform an arbitrary linear estimator into the one with a symmetric configuration, which will be proven to be an optimal estimator in the following sections.

Let a linear estimator be denoted by $\hat{f} = \hat{\theta} = \text{LN}(P, Q)$, with its underlying perturbation mechanism denoted by $M$. We construct a new perturbation mechanism $U$ as follows:

\begin{enumerate}
    \item Uniformly sample a permutation $\mathcal{X}_i$ from the space of all possible permutations of the dictionary $\mathcal{X} = [d]$.
    
    \item Permute the columns of the perturbation matrix $P$ according to $\mathcal{X}_i$ to derive a new perturbation matrix $P_i$, which defines a corresponding new perturbation mechanism $M_i$.
    
    \item Return the output $M_i(X)$ along with the chosen permutation $\mathcal{X}_i$.
\end{enumerate}

There are $D = d!$ possible permutations in total. Let $Z_i$ be a permutation matrix such that the operation $A Z_i$ reorders the columns of $A$ according to the permutation $\mathcal{X}_i$. Its transpose satisfies $Z_i^T Z_i = Z_i Z_i^T = I$. These permutation matrices possess the following property:
\begin{lemma}
For any $d \times d$ matrix $A$,
\begin{equation*}
    \frac{1}{D} \sum_{i=1}^{D} Z_i^T A Z_i = (a - b) I + b J,
\end{equation*}
where $I$ is an identity matrix and $J$ is an all-ones matrix. The coefficients are defined as $a = \frac{1}{d} \sum_{i=1}^{d} A_{ii}$ and $b = \frac{1}{d(d-1)} \sum_{i=1}^{d} \sum_{j \ne i, j=1}^{d} A_{ij}$.
\label{lem:perm_A}
\end{lemma}
\noindent Proof: The set of all permutations ensures that any index $i \in [d]$ is mapped to any $i' \in [d]$ with probability $\frac{1}{d}$. Furthermore, any pair of distinct indices $(i, j)$ with $i \ne j$ is mapped to all possible pairs of distinct values with a uniform probability of $\frac{1}{d(d-1)}$. $\qed$

Subsequently, we establish the following:
\begin{theorem}
    Perturbation $U$ has a symmetric configuration.
\label{thm:URP_sym}
\end{theorem}
\noindent Proof: For any permutation $i \in [D]$, the corresponding support probability matrix is given by $S_i^T P_i = Z_i^T S P Z_i$. Applying Lemma \ref{lem:perm_A}, we obtain:
\begin{equation*}
    \frac{1}{D} \sum_{i \in [D]} S_i^T P_i = (a - b) I + b J ,
\end{equation*}
\noindent where $a$ and $b$ are constants, aligned with Definition \ref{def:sym}. $\qed$
Meanwhile, an unbiased reconstruction matrix for $U$ can be derived from the original reconstruction matrix $Q$:
\begin{theorem}
    \begin{equation*}
    Q_{U} = 
    \begin{pmatrix}
    Q_1, Q_2, Q_3, ..., Q_D
    \end{pmatrix} 
\end{equation*}
\noindent is an unbiased reconstruction matrix for $U$, where $\forall i:Q_i = Z_i^T Q$.
\label{thm:perm_Q_unbiased}
\end{theorem}

\noindent Proof: Given $P_i = P Z_i$, it follows that $Q_i P_i = I$. Since each permutation is sampled uniformly at random, the perturbation matrix for $U$ is given by:
\begin{equation*}
    P_{U} = \frac{1}{D}
    \begin{pmatrix}
    P_1, P_2, P_3, ..., P_D
    \end{pmatrix}^T .
\end{equation*}

\noindent Consequently, we have
\begin{equation}
    Q_{U} P_{U} = \frac{1}{D} \sum_{i \in [D]} Q_i P_i = I . \qed \label{eq:perm_QP=I}
\end{equation}

As a result, an estimator based on a uniformly random permutation is formally defined as follows:
\begin{definition}
    $\text{URP}(P, Q) = \text{LN}(P_{U}, Q_{U})$, where 
    \begin{align*}
        & P_{U} = \frac{1}{D}(P Z_1, P Z_2, ..., P Z_D)^T \\
        & Q_{U} = (Z_1^T Q, Z_2^T Q, ..., Z_D^T Q)
    \end{align*}
    \noindent and $QP = I$.
\end{definition}

In analyzing its variance, following a derivation similar to Eq. \eqref{eq:perm_QP=I}, we obtain:
\begin{equation*}
    (Q_{U} \circ Q_{U}) P_{U} = \frac{1}{D} \sum_{i \in [D]} (Q_i \circ Q_i) P_i 
\end{equation*}

\noindent Applying Theorem \ref{thm:var_for_any_P}, we establish the following lemma:
\begin{lemma}
    \begin{align}
    \text{Var}(\hat{f}_{U} ) =  (Q_{U} \circ Q_{U}) P_{U}  \mathbf{f} - \mathbf{f} = \frac{1}{D}\sum_{i \in [D]} \text{Var}(\hat{f}_{i} ) ,  \label{eq:var_m_mean} \\
    \text{Var}(\hat{\theta}_{U} ) =  (Q_{U} \circ Q_{U}) P_{U}  \bm{\theta} - \bm{\theta}^2 = \frac{1}{D}\sum_{i \in [D]} \text{Var}(\hat{\theta}_{i} ) 
\end{align}
\noindent where $\hat{f}_{U} = \text{URP}(P, Q) $ and $\hat{f}_{i} = \hat{\theta}_{i} = \text{LN}(P_i, Q_i)$, respectively.
\label{lem:var_m_mean}
\end{lemma}

Substituting Lemma \ref{lem:var_m_mean} into Lemma \ref{lem:perm_A}, we obtain
\begin{multline*}
    \frac{1}{D} \sum_{i \in [D]} (Q_i \circ Q_i) P_i  = \frac{1}{D} \sum_{i \in [D]} Z_i^T (Q \circ Q) P Z_i = (\alpha - \beta) I + \beta J,
\end{multline*}
\noindent where $\alpha = \frac{1}{d} \sum_{i \in d} [(Q \circ Q) P)]_{ii}$ and $\beta = \frac{1}{d(d - 1)} \sum_{i \in d}  \sum_{j \ne i}   [(Q \circ Q) P)]_{ij}$. Consequently, we have proven that:

\begin{theorem}
    The variance of $\text{URP}(P, Q)$ satisfies
    \begin{align*}
        \text{Var}(\hat{f}_{U}) & = (\alpha - \beta)  \mathbf{f} + \beta - \mathbf{f}, \\
        \text{Var}(\hat{\theta}_{U}) & = (\alpha - \beta)  \bm{\theta} + \beta - \bm{\theta}^2,
    \end{align*}
\noindent where $\alpha = \frac{1}{d} \sum_{i \in [d]} [(Q \circ Q) P)]_{ii}$ \\ and $\beta = \frac{1}{d(d - 1)} \sum_{i \in [d]}  \sum_{j \ne i}   [(Q \circ Q) P)]_{ij}$. 
\label{thm:perm_var_aI+bJ}
\end{theorem}

Additionally, applying Theorem \ref{thm:perm_var_aI+bJ} to Theorem \ref{thm:l1_approx},  $\mathcal{L}_1$ can be formulated as a function of $\mathcal{L}_2$:
\begin{corollary}
For a frequency or distribution estimator $\hat{r}$ with a uniformly random permutation:
    \begin{equation*}
        \lim_{n \rightarrow \infty} \mathcal{L}_1(\hat{r}, n) = \sqrt{\frac{2d}{\pi} \mathcal{L}_2(\hat{r}, n)} .
    \end{equation*}
\label{cor:l1_to_l2}
\end{corollary}
\noindent The detailed proof is provided in Appendix 
\ref{sec:proof_thm:l1_approx}. 

\section{Strict Optimality in Precision}
\label{sec:strict_optimality}
This section frames the optimality of $\mathcal{L}_1$ and $\mathcal{L}_2$ as two distinct minimization problems:
\begin{equation}
    \min_{\hat{r}} \mathcal{L}_{1|2} (\hat{r}, n) = \min_P \min_{\hat{r} \in \mathcal{R}(P)} \mathcal{L}_{1|2} (\hat{r}, n)  , \label{eq:min_p_optimal}
\end{equation}
\noindent where $\hat{r}$ denotes either $\hat{f}$ or $\hat{\theta}$, and $\mathcal{R}(P)$ represents the set of all estimators that use matrix $P$ for perturbation ($\mathcal{R}$ stands for reconstruction). Subsequently, we define:
\begin{definition}
    An estimator $\hat{r} \in \mathcal{R}(P)$ is $P$-optimal if and only if $\forall P:\mathcal{L}_{1|2}(\hat{r}, n) = \min_{\hat{r} \in \mathcal{R}(P)} \mathcal{L}_{1|2} (\hat{r}, n) $.
\end{definition}

Sections \ref{sec:p_optimal_distribution} and \ref{sec:p_optimal_frequency} prove that a URP estimator can achieve $P$-optimality, while Section \ref{sec:derve_optimal} minimizes its $\mathcal{L}_1$ and $\mathcal{L}_2$ to establish optimal bounds.

\subsection{$P$-Optimal Distribution Estimator}
\label{sec:p_optimal_distribution}
We begin by analyzing $P$-optimality for distribution estimation. Since $\mathcal{L}_{1|2}$ is the supremum of $L_{1|2}$, it is naturally lower-bounded by any $L_{1|2}$:
\begin{equation*}
    \min_{\hat{\theta} \in \mathcal{R}(P)} \mathcal{L}_{1|2} (\hat{\theta}, n)  \ge \min_{\hat{\theta} \in \mathcal{R}(P)} L_{1|2}(\hat{\theta}, \bm{\theta}, n).
\end{equation*}
\noindent Subsequently, we prove that
\begin{theorem}
\begin{align*}
    & \min_{\hat{\theta} \in \mathcal{R}(P)} L_2(\hat{\theta}, \bm{\theta}, n)  \ge  \frac{1}{n} [\text{Tr}(\mathscr{I}_C^{+})]  \overset{\mathrm{def}}{=} L_2^{\downarrow}(P, \bm{\theta}, n). \\
    \lim_{n \rightarrow \infty} &  \min_{\hat{\theta} \in \mathcal{R}(P)} L_1(\hat{\theta}, \bm{\theta}, n)  \ge  \sum_{x \in [d]} \sqrt{\frac{2}{n \pi} (\mathscr{I}_C^{+})_{x,x}}   \overset{\mathrm{def}}{=} L_1^{\downarrow}(P, \bm{\theta}, n). 
\end{align*}
\noindent where $\mathscr{I}_C^{+} = (P^T \text{diag}(P \bm{\theta})^{-1} P)^{-1} + \bm{\theta} \bm{\theta}^T $ and $L_{1|2}^{\downarrow}$ represents the theoretical lower bound of $L_{1|2}$.
\label{thm:L2_fisher}
\end{theorem}
\noindent The proof, provided in Appendix \ref{sec:proof_thm:L2_fisher}, is inspired by the derivation of the Cramér–Rao bound using Fisher information. 

Consequently, We will establish the $P$-optimality of $\hat{\theta}_{U} = \text{URP}(P, Q)$ by proving:
\begin{equation}
    \min_Q \mathcal{L}_{1|2}(\text{URP}(P, Q), n) = L_{1|2}^{\downarrow}(P, \frac{1}{d} \mathbf{1}, n) 
    \label{eq:urp_p_optimal}
\end{equation}
\noindent where $\mathbf{1}$ denotes an all-ones column vector of size $d$. Note that the equality of $\mathcal{L}_1$ holds asymptotically as $n \to \infty$. Referring to Theorem \ref{thm:perm_var_aI+bJ}, we derive the $L_2$ norm of a URP estimator as:
\begin{multline}
    L_2(\hat{\theta}, \bm{\theta}, n)  = \sum_x (\alpha - \beta) \theta_x + \beta - \theta_x^2 =  \alpha + (d - 1) \beta - \bm{\theta}^T \bm{\theta} \\
    = \frac{1}{d} \sum_i [(Q \circ Q) P]_{ii} + \frac{1}{d} \sum_i \sum_{j \ne i} [(Q \circ Q) P]_{ij} - \bm{\theta}^T \bm{\theta} \\
    = \frac{1}{d} \sum_i \sum_j [(Q \circ Q) P]_{ij} - \bm{\theta}^T \bm{\theta} = \| Q \overline{P}^{\frac{1}{2}} \|_2 - \bm{\theta}^T \bm{\theta}. \label{eq:L2_to_QP_norm}
\end{multline}
\noindent where $\overline{P} = \text{diag}(\frac{1}{d} P \mathbf{1})$; $\|\cdot\|_2^2$ represents the squared $L_2$ norm of a matrix (i.e., the squared Frobenius norm). It follows that:
\begin{equation*}
    L_2(\hat{\theta}_U, \bm{\theta}, n)  \le L_2(\hat{\theta}_U, \frac{1}{d} \mathbf{1}, n) = \| Q \overline{P}^{\frac{1}{2}} \|_2 - \frac{1}{d} = \mathcal{L}_2(\hat{\theta}_U, n) .
\end{equation*}

Optimizing $\mathcal{L}_2$ is thus equivalent to minimizing $\| Q \overline{P}^{\frac{1}{2}} \|_2$. For the reconstruction matrix, we have:
\begin{theorem}
    Let $\hat{\theta}_{U} = \text{URP}(P, Q)$. For a fixed perturbation matrix $P$, the $\mathcal{L}_2$ loss is minimized when
    \begin{equation*}
        Q = (P^T  \overline{P}^{-1} P^T )^{-1} P^T \overline{P}^{-1} \overset{\mathrm{def}}{=} Q^*,
    \end{equation*}
    \noindent The resulting minimum $\mathcal{L}_2$ loss is given by
    \begin{equation*}
       \min_Q \mathcal{L}_2(\hat{\theta}_{U}, n)  = \frac{1}{n} [ \text{Tr}((P^T  \overline{P}^{-1} P^T )^{-1} ) - \frac{1}{d}],
    \end{equation*}
\noindent where $\text{Tr}$ denotes the trace of a matrix.
\label{thm:optimized_q}
\end{theorem}

\noindent Proof: Based on the Minimal Norm Least Squares Solution (MNLSS) \cite{campbell2009generalized}, for an underdetermined system $Ax = b$, the Moore–Penrose inverse $x = A^T(AA^T)^{-1}b$ minimizes $\|x\|_2$. Given the constraint $Q P = I$ (or equivalently $Q \overline{P}^{\frac{1}{2}} \overline{P}^{-\frac{1}{2}} P = I$), minimizing $\| Q \overline{P}^{\frac{1}{2}} \|_2$ via MNLSS yields
\begin{equation*}
    Q =  (P^T  \overline{P}^{-1} P^T )^{-1} P^T \overline{P}^{-1} .
\end{equation*}

\noindent The corresponding optimized $\mathcal{L}_2$ is derived as
\begin{multline*}
    \min_Q \mathcal{L}_2(\hat{\theta}_{U}, n)  \\ = \frac{1}{d} \sum_i \sum_j [(Q^* \circ Q^*) P]_{ij} - \frac{1}{d} =  \frac{1}{d} \sum_i  [(Q^* \circ Q^*) P \mathbf{1}]_{i} - \frac{1}{d} \\ =  \text{Tr}(Q^* \overline{P} (Q^*)^T) -\frac{1}{d}  = \text{Tr}((P^T  \overline{P}^{-1} P )^{-1}) - \frac{1}{d}. \qed
\end{multline*}
\noindent This proves the $\mathcal{L}_2$ component of Eq. \eqref{eq:urp_p_optimal}. Following Theorem \ref{thm:l1_approx}, we obtain:
\begin{equation}
     \lim_{n \rightarrow \infty} \mathcal{L}_1(\text{URP}(P, Q^*), n) = L_1^{\downarrow}(P, \frac{1}{d} \mathbf{1}, n) . \label{eq:urp_l1_optimal}
\end{equation}
\noindent  Consequently, we conclude:
\begin{theorem}
    $\text{URP}(P, Q^*)$ is a $P$-optimal distribution estimator.
\end{theorem}

\subsection{$P$-Optimal Frequency Estimator}
\label{sec:p_optimal_frequency}
We now extend the $P$-optimality established for distribution estimation to frequency estimation. First, consider the following sampling property:
\begin{theorem}
    Assume a dataset contains $n$ objects sampled i.i.d. from a categorical distribution $\mathcal{C}(\bm{\theta})$ and subsequently perturbed by an LDP mechanism. If the same unbiased estimator $\hat{f} = \hat{\theta}$ is employed for both frequency and distribution estimation, it satisfies:
    \begin{equation*}
        \sum_{\mathbf{f}} \Pr(\mathbf{f}  | \bm{\theta}) \text{Var}(\hat{f} | \mathbf{f}) = \text{Var}(\hat{\theta}) +  \frac{\bm{\theta}^2 - \bm{\theta}}{n} .
    \end{equation*}
\label{thm:var_theta_to_f}
\end{theorem}
\noindent The proof is provided in Appendix \ref{sec:proof_thm:var_theta_to_f}. By relating these variances into $L_2$ and applying Theorem \ref{thm:L2_fisher}, we obtain:
\begin{multline*}
    \sum_{\mathbf{f}} \Pr(\mathbf{f}  | \bm{\theta}) L_2(\hat{f}, \mathbf{f}, n) = L_2(\hat{\theta}, \bm{\theta}, n) +  \frac{\bm{\theta}^T \bm{\theta} - 1}{n} \\ \ge \frac{1}{n} [\text{Tr}((P^T \text{diag}(P \bm{\theta})^{-1} P)^{-1}) - 1].
\end{multline*}

Regarding frequency estimation, Eq. \eqref{eq:L2_to_QP_norm} is adjusted to $\mathcal{L}_2(\hat{f}_U) = \| Q \overline{P}^{\frac{1}{2}} \|_2 - 1$, which modifies Theorem \ref{thm:optimized_q} as follows:
\begin{equation*}
    \mathcal{L}_2(\hat{f}_U = \text{URP}(P, Q^*), n) = \frac{1}{n} [ \text{Tr}((P^T  \overline{P}^{-1} P )^{-1}) - 1], 
\end{equation*}
\noindent where $\overline{P} = \text{diag}(P \frac{1}{d} \mathbf{1})$. Since the worst-case loss $\mathcal{L}_2$ is lower-bounded by the average $L_2$ loss, we have:
\begin{multline*}
    \min_{\hat{f} \in \mathcal{R}(P)} \mathcal{L}_2 (\hat{f}, n) \ge \min_{\hat{f} \in \mathcal{R}(P)} \sum_{\mathbf{f}} \Pr(\mathbf{f} | \bm{\theta} = \frac{1}{d} \mathbf{1}) L_2(\hat{f}, \mathbf{f}, n) \\ \ge \frac{1}{n} [ \text{Tr}((P^T  \overline{P}^{-1} P )^{-1}) - 1] = \mathcal{L}_2(\text{URP}(P, Q^*), n),
\end{multline*}
\noindent By definition, $\mathcal{L}_2(\text{URP}(P, Q^*), n) \ge \min_{\hat{f} \in \mathcal{R}(P)} \mathcal{L}_2 (\hat{f}, n)$, implying that these quantities must be identical. Thus, $\text{URP}(P, Q^*)$ is also $P$-optimal for frequency estimation. Similarly to Eq. \eqref{eq:urp_l1_optimal}, this $P$-optimality extends to $\mathcal{L}_1$. That is,
\begin{theorem}
    $\text{URP}(P, Q^*)$ is a $P$-optimal frequency estimator.
\end{theorem}

\subsection{Strict Lower Bound}
\label{sec:derve_optimal}
Since $\text{URP}(P, Q^*)$ is $P$-optimal for both frequency and distribution estimation, Eq. \eqref{eq:min_p_optimal} ensures that $\min_P \mathcal{L}_{1|2} (\text{URP}(P, Q^*), n)$ constitutes the strict lower bound for $\mathcal{L}_{1|2}$. This section derives these values as follows:
\begin{enumerate}
    \item Among all perturbation matrices sharing the same self-support probability, determine the optimal $P^*$ that minimizes $\|Q^* \overline{P}^{\frac{1}{2}}\|_2$.
    \item Prove that $\text{URP}(P^*, Q^*)$ achieves an Optimal Symmetric Configuration (OSC, as per Definition \ref{def:osc}).
    \item Derive the $\mathcal{L}_2$ loss of the OSC as a function of dictionary size $d$, privacy parameter $\epsilon$, dataset size $n$, and support size $k$.
    \item Minimize the $\mathcal{L}_2$ loss of the OSC with respect to the support size $k$ to establish the strict lower bound for all estimators.
\end{enumerate}
These steps are formally expressed by the following sequence of inequalities:
\begin{multline*}
    \forall P: \mathcal{L}_2(\text{URP}(P, Q^*)) 
    \ge \mathcal{L}_2(\text{URP}(P^*, Q^*)) \\ = \mathcal{L}_2(\text{OSC}(d, \epsilon, n, k)) \ge \min_k \mathcal{L}_2(\text{OSC}(d, \epsilon, n, k)) .
\end{multline*}

Notably, this section focuses on \textbf{frequency estimation}, but all results are applicable to distribution estimation by adjusting the last term of their variance expressions, i.e., by replacing $\mathbf{f}$ with $\bm{\theta}^2$. Consequently, $\mathcal{L}_2(\hat{\theta}, n) = \mathcal{L}_2(\hat{f}, n) + \frac{1 - 1/d}{n}$.

We now proceed to find the optimal $P$ to minimize $\mathcal{L}_2$. Recall that Lemma \ref{lem:constant_sp} establishes that an extremal configuration is characterized by a constant self-support probability $p^*$, such that $\forall x: \sum_o e^{\epsilon} S_{o, x} p_o = p^*$. Building on this, we provide the following result:

\begin{theorem}
    Among all perturbation matrices with an extremal configuration and a fixed self-support probability $p^*$, the one with a constant support size minimizes the $\mathcal{L}_2$ loss of $\text{URP}(P, Q^*)$.
\label{thm:constant_support_optimality}
\end{theorem}

\noindent The detailed proof is provided in Appendix \ref{sec:proof_thm:constant_support_optimality}. The core idea involves relating $\mathcal{L}_2$ to the eigenvalues of $P^T \overline{P}^{-1} P$, denoted by $\lambda_1, \lambda_2, \dots, \lambda_d$. The proof demonstrates that $\lambda_1 = d$ always holds. Letting $\Lambda = \sum_{i=2}^d \lambda_i$, we have:

\begin{equation*}
    \text{Tr}(P^T \overline{P}^{-1} P)  = \frac{1}{d} + \sum_{i=2}^d \frac{1}{\lambda_i} \ge \frac{1}{d} + \frac{(d-1)^2}{\Lambda} \ge \frac{1}{d} + \frac{(d-1)^2}{\max \Lambda} .
\end{equation*}

\noindent Appendix \ref{sec:proof_thm:constant_support_optimality} further shows that when the support size for all responses is constant (i.e., $\forall o: k_o = k$), both equalities are achieved simultaneously, thereby minimizing $\mathcal{L}_2$. $\qed$

\begin{theorem}
If a perturbation matrix $P^*$ has an extremal configuration and constant support size, the reconstruction of $\text{URP}(P^*, Q^*)$ is equivalent to Eq. \eqref{eq:sym_fx}.
\label{thm:optimized_urp_reconstruction}
\end{theorem}

\noindent The detailed proof is provided in Appendix \ref{sec:proof_thm:optimized_urp_reconstruction}. Intuitively, given a constant support size, the reconstruction matrix takes the form $Q = A [(e^{\epsilon} - 1) S^T + J^T ]$, where $A$ is a $d \times d$ matrix. Here, $S^T$ maps the responses to their supported original values, and $J^T$ simply adds a constant offset. Consequently, the reconstruction is equivalent to a linear transformation of the frequencies of the values supported by the responses, which matches the operation in Eq. \eqref{eq:sym_fx}. $\qed$

\begin{definition}
    An estimator has an Optimal Symmetric Configuration (OSC) if it has an extremal and symmetric configuration with constant support, and employs Eq. \eqref{eq:sym_fx} to reconstruct the frequencies or distributions.
\label{def:osc}
\end{definition}

Theorems \ref{thm:constant_support_optimality}, \ref{thm:optimized_urp_reconstruction}, and \ref{thm:URP_sym} establish that a URP estimator achieves an OSC. Thus, we have:
\begin{equation*}
  \min_{\hat{f}_{\text{OSC}}}   \mathcal{L}_2(\hat{f}_{\text{OSC}}, n) \le \min_{\hat{f}_{\text{URP}}}   \mathcal{L}_2(\hat{f}_{\text{URP}}, n) .
\end{equation*}

\noindent Furthermore, since $\min_{\hat{f}_{\text{URP}}} \mathcal{L}_2(\hat{f}_{\text{URP}})$ has been established as an optimal bound, it follows that $\min_{\hat{f}_{\text{OSC}}} \mathcal{L}_2(\hat{f}_{\text{OSC}})$ similarly constitutes an optimal bound.

To derive $\mathcal{L}_2$ of an OSC estimator, we first formulate the variance of Eq. \eqref{eq:sym_fx} as a function of the dictionary size $d$, privacy parameter $\epsilon$, dataset size $n$, and support size $k$. The extremal configuration ensures that $k$ can be directly derived from $p^*$, as shown in Lemma \ref{lem:constant_sp}. Furthermore, the cross-support probability $q^*$ of OSC is derived as follows:
\begin{lemma}
An optimal symmetric configuration satisfies
\label{lem:p_q_star_k}
    \begin{align*}
    & p^* = \frac{e^{\epsilon} k}{k(e^{\epsilon} - 1) + d} \\
    & q^* =  \frac{(e^{\epsilon} - 1) k (k - 1)}{(d-1) [k (e^{\epsilon} - 1) + d] }  + \frac{k}{k (e^{\epsilon} - 1) + d}
    \end{align*}
\end{lemma}

\noindent This proof is presented in Appendix \ref{sec:proof_lem:p_q_star_k}. Subsequently, given Theorem \ref{thm:sym_exp_var}, the variance of an OSC frequency estimator is derived as
\begin{lemma}
\label{lem:var_lb}
    \begin{multline}
        \text{Var}(\hat{f}_{\text{OSC}}(x)) = \frac{ [k (e^{\epsilon} - 1) + d -1 ] [k (e^{\epsilon} - 1) + d - e^{\epsilon} ]  }{ n k (d - k)  (e^{\epsilon} - 1)^2}  \\
        + f(x) \frac{ k^2(e^{\epsilon} - 1) + k (2d - e^{\epsilon} - 1) + d - d^2 }{ n k (d - k)  (e^{\epsilon} - 1)} \label{eq:var_fx_as_k}
    \end{multline}
\end{lemma}

\noindent By substituting $\text{Var}(\hat{f}_{\text{OSC}}(x))$ into $\mathcal{L}_1$ and $\mathcal{L}_2$, we optimize $k$ to minimize these losses:
\begin{theorem}
When $k = \frac{d}{e^{\epsilon} + 1} \ge 1$, $\mathcal{L}_1$ and $\mathcal{L}_2$ achieve their minimum values:
    \begin{align}
        & \mathcal{L}_1^*(n) = \min_{k} \mathcal{L}_1(\hat{f}, n) \approx \frac{ \sqrt{ 2 (d - 1) [4 d e^{\epsilon} - (e^{\epsilon} + 1)^2  }] }{ \pi \sqrt{n} (e^{\epsilon} - 1)} .\\
        & \mathcal{L}_2^*(n) = \min_{k} \mathcal{L}_2(\hat{f}, n) = \frac{(d - 1) [4 d e^{\epsilon} - (e^{\epsilon} + 1)^2]}{n d (e^{\epsilon} - 1)^2} 
    \end{align}
\label{thm:optimized_l1_l2}
\end{theorem}

\noindent The detailed proof is provided in Appendix \ref{sec:proof_thm:optimized_l1_l2}. The core methodology involves finding the minima through differentiation with respect to $k$. Consequently, the derivation throughout this section establishes a sufficient condition for an optimal estimator:
\begin{corollary}
    An estimator is $(\mathcal{L}_1, \mathcal{L}_2)$-optimal if it has an optimal symmetric configuration with $k = \frac{d}{e^{\epsilon} + 1}$. 
\end{corollary}

The optimality established in Theorem \ref{thm:optimized_l1_l2} assumes $k$ to be an integer. When the optimal $k$ is non-integral, one may select the nearest integers that provide the higher precision. In such cases, the deviation is bounded as follows:
\begin{corollary}
    When $k$ is a nearby integer of $\frac{d}{e^{\epsilon} + 1}$, we have
    \begin{align*}
    & \min_{\hat{f}_{OSC}} \mathcal{L}_1(\hat{f}_{OSC}, n) \le \mathcal{L}_1^*(n) \sqrt{1 + \alpha} , \\
    & \min_{\hat{f}_{OSC}} \mathcal{L}_2(\hat{f}_{OSC}, n) \le \mathcal{L}_2^*(n) (1 + \alpha) ,
    \end{align*}
\noindent where $\alpha = \frac{(d - 1) (e^{\epsilon} + 1)^4}{(2d + e^{\epsilon} + 1) (2d e^{\epsilon} - e^{\epsilon} - 1) (4d e^{\epsilon} - (e^{\epsilon} + 1)^2)}$ .
\label{cor:l1_l2_deviation}
\end{corollary}
\noindent The proof is provided in Appendix \ref{sec:proof_cor:l1_l2_deviation}.

Notably, when $e^{\epsilon} + 1 \ge d$, the support size $k$ cannot be less than one. By setting $k = 1$ in this regime, the optimized $\mathcal{L}_1$ and $\mathcal{L}_2$ losses are given by:

\begin{corollary}
When $e^{\epsilon} + 1 \ge d$, the minimum values for $\mathcal{L}_1$ and $\mathcal{L}_2$ are:
    \begin{align}
        & \min_{\hat{f}} \mathcal{L}_1 (\hat{f}, n) \approx \frac{ \sqrt{ 2 d (d - 1) (d + 2 e^{\epsilon} - 2)}  }{ \pi\sqrt{n} (e^{\epsilon} - 1 ) }  .\\
        & \min_{\hat{f}} \mathcal{L}_2 (\hat{f}, n) = \frac{ (d - 1) (d + 2 e^{\epsilon} - 2) }{ n (e^{\epsilon} - 1)^2 } 
    \end{align}
\label{cor:l1_l2_k=1}
\end{corollary}

\noindent This result is established by expressing $\mathcal{L}_1$ and $\mathcal{L}_2$ as functions of $k$ and substituting $k = 1$. Further details are available in Appendix \ref{sec:proof_cor:l1_l2_k}.

\section{Communication Cost}
This section evaluates the upper bound of the communication cost for an optimal frequency or distribution estimator. Let $\mathcal{O}$ represent all potential responses from a LDP client, and $p_o$ denote the base probability of a given response $o \in \mathcal{O}$. To prevent privacy leakage through the encoding process, we require all responses to utilize the same encoding scheme, regardless of the original value that generated them. Therefore, responses are encoded based on their prior distribution. Shannon's Source Coding Theorem \cite{shannon1948mathematical} establishes that if the prior distribution of $\mathcal{O}$ is known for any response $o$ as $\rho_o$, the optimal communication cost for $\mathcal{O}$ is $-\sum_o \rho_o \log_2(\rho_o)$ bits. In scenarios where the prior distribution is unknown, the upper bound of the optimal communication cost is $\log_2(|\mathcal{O}|)$ bits, which occurs when $\forall o: \rho_o = \frac{1}{|\mathcal{O}|}$, i.e., the logarithm of the number of possible responses.

\begin{theorem}
    The communication cost of an optimal estimator can be as low as $\log_2(\frac{d(d-1)}{2} +1)$ bits. 
    \label{thm:communication_cost_ub}
\end{theorem}

The detailed proof is provided in Appendix \ref{sec:proof_communication_cost_ub}. The core idea is to demonstrate that at most $\frac{d(d-1)}{2} + 1$ responses are necessary to construct an optimal estimator with an optimal symmetric configuration (Definition \ref{def:osc}). To derive this bound, we prove that satisfying a symmetric configuration requires that $\forall i < j: \sum_o S_{o, i} S_{o, j} p_o = \frac{k(k-1)}{(d-1)[k(e^{\epsilon}-1) + d]}$ holds true. This corresponds to $\frac{d(d-1)}{2}$ distinct constraints. By leveraging Carathéodory's theorem \cite{barvinok2025course}, we establish that only $\frac{d(d-1)}{2} + 1$ responses are needed; thus, the upper bound of the communication cost is $\log_2\left(\frac{d(d-1)}{2} + 1\right)$ bits.

There are special cases where the communication cost can be further reduced to $\log_2 d$. For example, when $d = 2^i$ for $i \in \mathbb{N}^+$ and $k = \frac{d}{2}$ (i.e., as $\epsilon \rightarrow 0$), the support matrix $S$ can be constructed as a Hadamard matrix with $d$ rows, each maintaining a constant support size of $\frac{d}{2}$. This construction is equivalent to the Hadamard Response \cite{acharya2019hr}. Furthermore, assuming the base probability $p_o$ is constant across all responses, the construction of $S$ is equivalent to a Balanced Incomplete Block Design (BIBD). For more details, refer to \cite{bose1939construction}. Depending on the specific values of $(d, k)$, there may exist BIBDs that require only $d$ responses \cite{park2024exactly}.

Although it is always preferable to identify the optimal communication cost, i.e., a tight lower bound, which this paper does not derive, a communication cost of $O(\log d)$ bits remains practically feasible even for a very large dictionary size $d$.

\section{Optimal Algorithms}
\label{sec:optimized_algorithms}
\subsection{Subset Selection}

Subset Selection, originally proposed by \cite{ye2018subset}, returns a random set of $k$ distinct elements from the dictionary $\mathcal{X} = [d]$. The output space $\mathcal{O}$ consists of all $\binom{d}{k}$ possible combinations, where each response $o \in \mathcal{O}$ shares identical base probability $p_o$ (see Definition \ref{def:extremal}). If a subset contains (supports) the input value $x$, the probability of Subset Selection generating that response is $e^{\epsilon}$ times greater than the probability of generating a response that does not contain $x$.

This mechanism naturally follows an optimal symmetric configuration. Consequently, by applying the optimized $k$ from Theorem \ref{thm:optimized_l1_l2}, Subset Selection serves as an optimal estimator. Given that it possesses $\binom{d}{k}$ possible responses with identical base probabilities, the communication cost is bounded by $k \log_2 \left( \frac{e d}{k} \right)$ bits, following the inequality $\binom{d}{k} \le \left( \frac{e d}{k} \right)^k$.

\subsection{Optimized Count Mean Sketch}
\label{sec:ocms}
Count-Mean Sketch (CMS), originally proposed by \cite{apple_privacy}, was further refined by \cite{pan2025improving} to achieve state-of-the-art precision while maintaining a logarithmic communication cost. The simplified workflow for CMS is as follows:

\begin{enumerate}
    \item Determine the hash range $B$ (the number of possible hashed values) based on the privacy parameter $\epsilon$.
    \item Uniformly sample a hash function $h$ from a hash family $\mathcal{H}$, and map an object's original value $x$ to a hashed value $y = h(x)$.
    \item Apply a randomized response mechanism to perturb $y$ within the range $\{1, 2, \dots, B\}$, generating a perturbed value $z$. The tuple $(h, z)$ is then sent to the server.
    \item Frequencies or distributions are estimated based on the pairs $(h, z)$ collected from all objects, using a formula equivalent to Eq. \eqref{eq:sym_fx} as provided by \cite{pan2025improving}.
\end{enumerate}

This section demonstrates that CMS is a nearly optimal frequency or distribution estimator, subject to the following minor modifications: (1) expanding the original dictionary size $d$ to $d'$, where $d'$ is the smallest prime number such that $d' \ge d$; (2) set hash range $m = \text{round}(1 + e^{\epsilon})$ and (3) constructing the hash family as:
\begin{multline*}
    \mathcal{H} = \{ h(x) = ((a x + b) \mod {d'}) \mod B | \\
     a \in [1, d'-1], b \in [0, d'-1] \} .
\end{multline*}

This construction differs from \cite{pan2025improving} by explicitly disallowing $a = 0$. This modification improves the collision probability from approximately $\frac{1}{B^2}$ to approximately $\frac{d' - B}{B^2(d' - 1)}$ (see Appendix \ref{sec:proof_thm:cms_equivalance} for details). We refer to this variant as Optimized Count Mean Sketch (OCMS).

First, we demonstrate that OCMS is equivalent to sampling from two optimal symmetric configurations:
\begin{theorem}
The output distribution of OCMS with dictionary size $d'$ and $B$ buckets is equivalent to sampling between two optimal symmetric configurations:
\begin{itemize}
    \item A configuration with support size $\lfloor \frac{d'}{B} \rfloor + 1$, sampled with probability $p_{\alpha} = \frac{d' \mod B}{B}$, and
    \item A configuration with support size $\lfloor \frac{d'}{B} \rfloor$, sampled with probability $1 - p_{\alpha}$.
\end{itemize}
\label{thm:cms_equivalance}
\end{theorem}

\noindent The detailed proof is provided in Appendix \ref{sec:proof_ocms}. The core intuition is that any response $(h, z)$ supports all elements mapped to bucket $z$. Each bucket contains either $\lfloor \frac{d'}{B} \rfloor + 1$ or $\lfloor \frac{d'}{B} \rfloor$ elements. By grouping all possible responses according to their support size, each group constitutes a distinct configuration; thus, OCMS can be reconstructed as a weighted sampling of these two configurations.

This leads to the following result:
\begin{corollary}
    The variance of OCMS is given by
    \begin{multline*}
        \text{Var}(\hat{f}_{OCMS}(x)) = p_{\alpha} Var(\hat{f}^*(x | k= \lfloor \frac{d'}{B} \rfloor +1)) + \\ (1 -  p_{\alpha}) Var(\hat{f}^*(x | k= \lfloor \frac{d'}{B} \rfloor)) ,
    \end{multline*}
\noindent where $\hat{f}^*(x|k)$ represents the estimator with an optimal symmetric configuration and support size $k$.
\label{cor:cms_var}
\end{corollary}

\noindent The full proof is provided in Appendix \ref{sec:proof_ocms}. Subsequently, we demonstrate that OCMS approaches optimal precision:
\begin{corollary}
    Define $\alpha = \frac{(d-1) (e^{\epsilon} + 1)^4}{(4d e^{\epsilon} - (e^{\epsilon } + 1)^2) (d - e^{\epsilon} - 1) (d e^{\epsilon} + e^{\epsilon} + 1)}$, $\beta = \frac{(d' - d) (e^{\epsilon} + 1)^2 (2d d' - d - d')}{d'^2 (d - 1) (4de^{\epsilon} - (e^{\epsilon} + 1)^2)} $. If OCMS has an optimized hash range $B = \text{round}(1 + e^{\epsilon})$ , then
    \begin{align*}
    & \mathcal{L}_{1, OCMS}^*(n) \le \mathcal{L}_{1}^*(n)  \sqrt{(1 + \alpha) (1 + \beta)} \\
    & \mathcal{L}_{2, OCMS}^*(n)  \le \mathcal{L}_{2}^*(n)  (1 + \alpha) (1 + \beta)
    \end{align*}
    \noindent where $\mathcal{L}_{1}^*$ and $\mathcal{L}_{2}^*$ represent the optimized $\mathcal{L}_1$  and $\mathcal{L}_2$ given dictionary size $d$ and privacy parameter $\epsilon$, respectively.
    \label{cor:cms_deviation}
\end{corollary}

\noindent The proof is detailed in Appendix \ref{sec:proof_ocms}. Notably, when $B = \text{round}(1 + e^{\epsilon})$, OCMS is equivalent to sampling from two optimal symmetric configurations with support sizes of $\lfloor \frac{d'}{1 + e^{\epsilon}} \rfloor$ and $\lceil \frac{d'}{1 + e^{\epsilon}} \rceil$, respectively. In the regime where $d \gg e^{\epsilon}$, both $\alpha$ and $\beta$ vanish toward zero. For instance, with $\epsilon = 1$, $d = 100$, and $d' = 101$, the $\mathcal{L}_{2}$ error of OCMS is only $0.09\%$ higher than the theoretical lower bound.

\subsection{Weighted Subset Selection}

Theorem \ref{thm:communication_cost_ub} proves that at most $\frac{d(d-1)}{2} + 1$ responses are required to construct an optimal estimator. These responses constitute a subset of $S^{\star}$, which denotes the set of all possible $\binom{d}{k}$ combinations. We can develop an algorithm to identify these $\frac{d(d-1)}{2} + 1$ responses and construct the corresponding matrices $S$ and $P$. The core construction procedure is as follows:

\begin{enumerate}
    \item Sample $m$ responses from $S^{\star}$ either uniformly or via a heuristic, where $m \ge \frac{d(d-1)}{2} + 1$.
    \item For each sampled response $o$, compute every $S_{o,i} S_{o,j}$ for $i < j$ as a vector $v_o$.
    \item Construct a matrix $A = [v_1, \dots, v_m]$ and a vector $y$ of size $\frac{d(d-1)}{2}$, where all elements are set to $\frac{k(k-1)}{(k(e^{\epsilon} - 1) + d) (d - 1)}$, i.e., the expected value of $\sum_o p_o S_{o, i} S_{o, j}$.
    \item Find a solution $\mathbf{p}$ for the system $A \mathbf{p} = y$, subject to $\mathbf{p} \ge 0$. This can be achieved using Linear Programming (LP) or Non-Negative Least Squares (NNLS). These solvers typically yield sparse solutions, generally resulting in no more than $\frac{d(d-1)}{2} + 1$ positive values. Refer to the proof of Theorem \ref{thm:communication_cost_ub} for why solving $\mathbf{p}$ for $A \mathbf{p} = y$.
    \item If a solution is successfully found, let $\mathbf{p} = [p_1, p_2, \dots, p_m]$ and construct $S$ by stacking the support vectors of all responses where $p_o > 0$.
    \item If the system is infeasible, return to step (1) and resample.
\end{enumerate}

This algorithm is referred to as Weighted Subset Selection (WSS). The heuristic in Step (1) can be implemented as follows:

\begin{enumerate}
    \item \textbf{Track Support}: Monitor the frequency with which each original value $x$ is supported by existing sampled responses using $\sum_o S'_{o, x}$, where $S'$ denotes the support matrix of the already sampled responses.
    \item \textbf{Weighted Sampling}: Randomly select $k$ values using weights proportional to $\exp\left(-\sum_o S'_{o, x}\right)$. This ensures that values with lower current support are assigned a higher probability of being sampled in the next iteration.
\end{enumerate}

In our experiments, this heuristic typically generates a valid $S$ when $m = d^2$. The constraint $A \mathbf{p} = y$ inherently guarantees an optimal symmetric configuration (see the proof of Theorem \ref{thm:communication_cost_ub}). Consequently, it serves as an optimal estimator provided that the support size $k$ is optimized. Since LP or NNLS solvers tend to produce sparse solutions, we observe that $\mathbf{p}$ generally contains no more than $\frac{d(d-1)}{2} + 1$ positive values. Thus, the communication cost is typically bounded by $\log_2\left(\frac{d(d-1)}{2} + 1\right)$. If a specific solution for $\mathbf{p}$ contains more responses than this bound, the process can be restarted to find a sparser alternative.

Notably, LP or NNLS solvers have a computational complexity of $O(H^2W)$, where $H$ and $W$ represent the height and width of $A$, respectively. In this context, $H = \frac{d(d-1)}{2}$ and $W = m = O(d^2)$, resulting in a total computational complexity of $O(d^6)$. A practical approach to adopting WSS is to precompute valid support matrices $S$ for common values of $d$ and $k$; clients can then simply download and reuse these matrices for each round of frequency or distribution estimation. Furthermore, as the proof of Corollary \ref{cor:cms_deviation} demonstrates that extending a dictionary to a larger size has a negligible impact on precision, one can simply use the next available precomputed $d$ for which a valid $S$ exists.

\subsection{Choosing an Optimal Algorithm}
A comparison of the three aforementioned algorithms is presented below:
\begin{table}[ht]
\centering
\begin{threeparttable}
\begin{tabular}{| c | c | c | c |}

\hline
 & SS & OCMS & WSS \\
 \hline
 Precision & \cellcolor{green!50} $\mathcal{L}^*$ & \cellcolor{green!20} $\mathcal{L}^* (1 + O(\frac{1}{d}))$ & \cellcolor{green!50} $\mathcal{L}^*$  \\
  \hline
 \makecell{Communication \\ Cost} & \cellcolor{red!50} $ O(\frac{d}{e^{\epsilon} }) $ & \cellcolor{green!20}  $ 2 \epsilon \log_2 (2d)  $ &  \cellcolor{green!50} \makecell{ $ 2 \log_2(d) $ } \\
 \hline
  \makecell{Precomputation \\ Cost} & 0 & 0 &  \cellcolor{red!50}  $O(d^6)$ \\
\hline

\hline
\end{tabular}
\begin{tablenotes}
    \item Note: $\mathcal{L}^*$ can be either $\mathcal{L}_1$ or $\mathcal{L}_2$. 
    \item Abbreviations: SS - Subset Selection, OCMS - Optimized Count Mean Sketch, WSS - Weighted Subset Selection.
\end{tablenotes}
\label{table:comparison}
\end{threeparttable}
\end{table}

When selecting an optimal frequency or distribution estimator, the choice largely depends on the dictionary size $d$. If $d$ is small, WSS or SS are viable options; otherwise, OCMS is preferable due to its low communication and precomputation overhead. To determine whether a dictionary is sufficiently large for OCMS to be effective, one may refer to Corollary \ref{cor:cms_deviation}. For example, when $\epsilon = 1$ and $d \ge 100$, $\mathcal{L}_2$ of OCMS is at most 0.09\% above the optimal loss, a gap that continues to narrow as $d$ increases.

Additionally, if the parameters of WSS have been precomputed for a specific privacy parameter $\epsilon$ and a dictionary size larger than your own, you may extend your dictionary to leverage these precomputed values.

\begin{figure*}[ht!]
\centering
\subfloat[]{\includegraphics[width=0.32\textwidth]{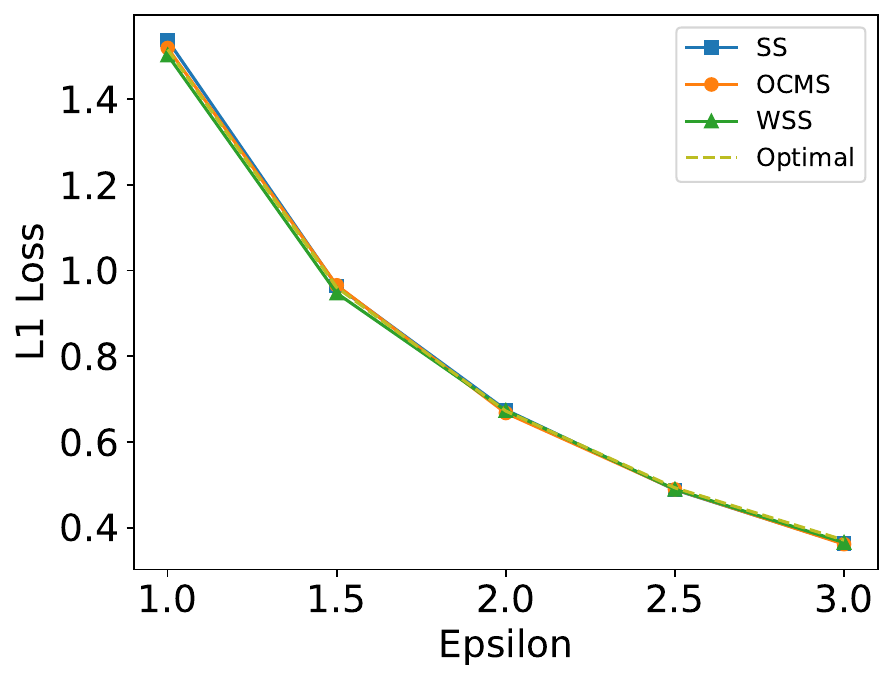}}
\subfloat[]{\includegraphics[width=0.32\textwidth]{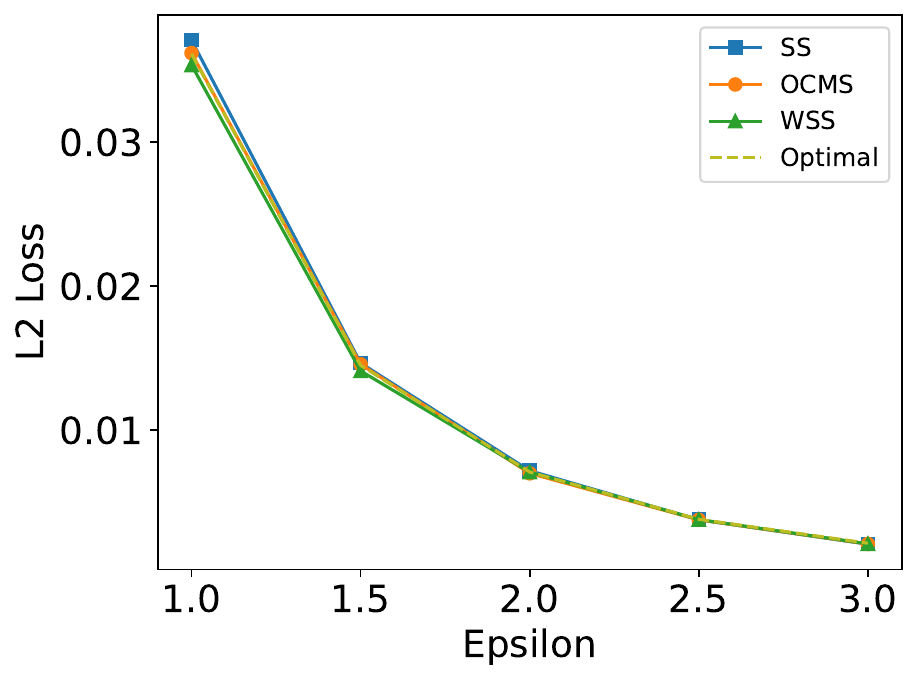}}%

\caption{$\mathcal{L}_1$ and $\mathcal{L}_2$ losses vs. privacy parameter $\epsilon$ given the Zipf distribution. See Section \ref{sec:zipf} for details.}
\label{fig:zipf} 

\subfloat[]{\includegraphics[width=0.32\textwidth]{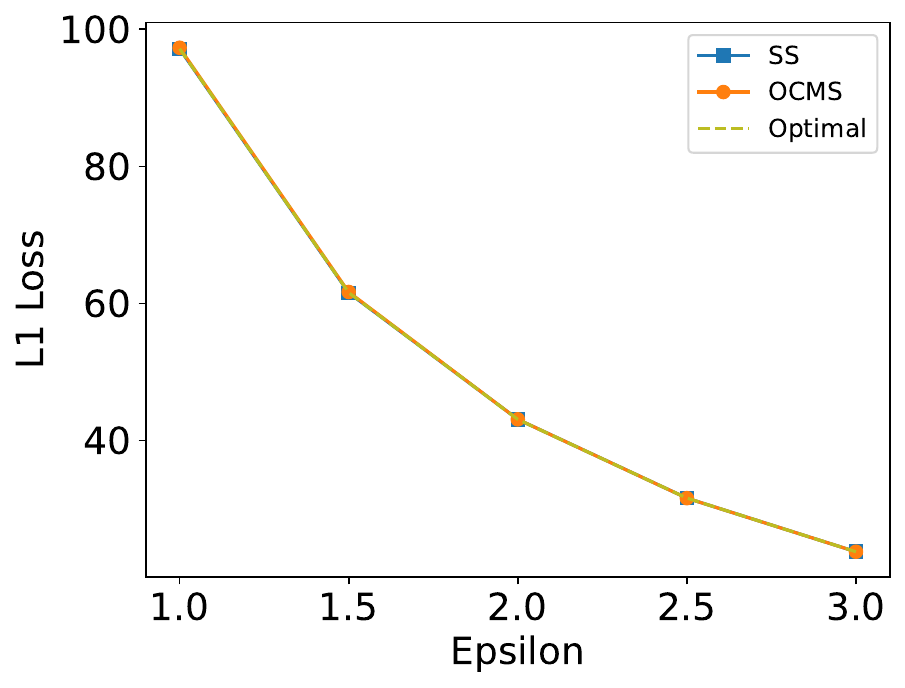}}
\subfloat[]{\includegraphics[width=0.32\textwidth]{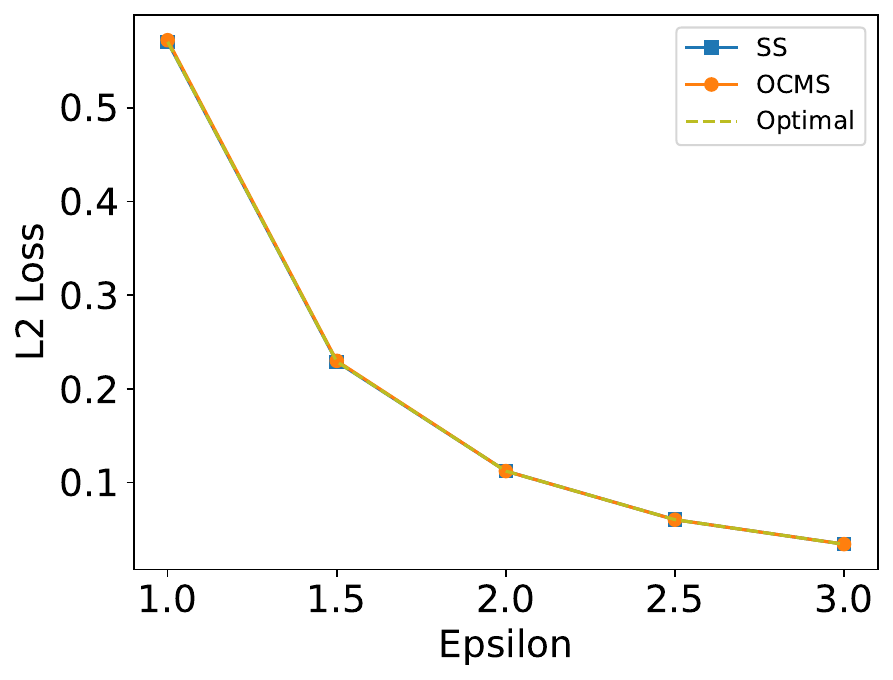}}%

\caption{$\mathcal{L}_1$ and $\mathcal{L}_2$ losses vs. privacy parameter $\epsilon$ given the mini Kosarak dataset. See Section \ref{sec:kosarak} for details.}
\label{fig:kosarak} 
\end{figure*}

\section{Experiment}
\label{sec:experiment}
This section empirically evaluates if the aforementioned optimal algorithms achieve the theoretical $\mathcal{L}_1$ and $\mathcal{L}_2$ losses. We conduct two experiments, each consisting of 100 independent runs. The $\mathcal{L}_1$ and $\mathcal{L}_2$ losses of the evaluated algorithms are measured as:
\begin{align*}
    & L_1 = \frac{1}{t} \sum_t \sum_{x \in [d]} |\hat{r}(x) - r(x)| \\
    & L_2 = \frac{1}{t} \sum_t \sum_{x \in [d]} (\hat{r}(x) - r(x))^2 
\end{align*}
\noindent where $t$ represents the number of experiment runs, $r$ can be either frequency $f$ or distribution $\theta$. Since $\mathbb{E}[L_1] = \mathcal{L}_1$ and $\mathbb{E}[L_2] = \mathcal{L}_2$, these measures are unbiased and should closely approximate the theoretical $\mathcal{L}_1$ and $\mathcal{L}_2$ when $t$ is large enough.

\subsection{Zipf Distribution}
\label{sec:zipf}
The first experiment samples datasets from a specific Zipf distribution to estimate the distribution $\bm{\theta}$, which is defined as:
\begin{equation*}
    \forall x\in [d]: \theta(x) = \text{Zipf}(x) \propto \frac{1}{x^2} ,
\end{equation*}

\noindent The dictionary size $d$ is set to 100, and 10,000 objects are sampled from this distribution. All three optimal algorithms introduced in Section \ref{sec:optimized_algorithms} are evaluated for their $\mathcal{L}_1$ and $\mathcal{L}_2$. The results are illustrated in Fig. \ref{fig:zipf}.

The results show that all three algorithms are practically indistinguishable from one another and align perfectly with the optimal $\mathcal{L}_1$ and $\mathcal{L}_2$ curves. This empirical evidence supports our theory that both Subset Selection (SS) and Weighted Subset Selection (WSS) are optimal. Furthermore, it validates Corollary \ref{cor:cms_deviation}: when the dictionary size is sufficiently large ($d = 100$ in this case), the Optimized Count Mean Sketch (OCMS) performs indistinguishably from an optimal algorithm.

\subsection{Real-world Dataset}
\label{sec:kosarak}
The second experiment estimates the frequency of Kosarak \cite{kosarak}, a real-world dataset containing click-stream data from a Hungarian online news portal. For our experiment, we reduce this dataset to 1\% of its original size (denoted as "mini Kosarak"), which still contains over 170,000 objects with a dictionary size $d = 26,000$. In this setup, we only evaluate Subset Selection (SS) and Optimized Count Mean Sketch (OCMS). Weighted Subset Selection (WSS) is omitted due to its high precomputation cost for a dictionary of this size. The results are plotted in Fig. \ref{fig:kosarak}.

Both SS and OCMS align perfectly with the optimal $\mathcal{L}_1$ and $\mathcal{L}_2$ curves. These findings further corroborate that SS is optimal and that OCMS is equivalent to an optimal algorithm for large dictionary sizes, as proven by Corollary \ref{cor:cms_deviation}.

\section{Conclusion}
This paper establishes the strict optimality of frequency and distribution estimation under LDP by proving that the optimal symmetric configuration achieves the theoretical lower bound of $\mathcal{L}_2$ loss. The theoretical $\mathcal{L}_1$ lower bound is also achieved asymptotically. We further demonstrate that the communication cost of an optimal estimator can be reduced to $\log_2 \left( \frac{d(d-1)}{2} + 1 \right)$ bits, where $d$ is the dictionary size, and we develop an algorithm to construct an optimal estimator within this communication bound. Consequently, our results offer a practical deployment strategy: Optimized Count-Mean Sketch is recommended for large dictionaries, while weighted and original Subset Selection are ideal for smaller scales. Experimental evaluations confirm that these optimal algorithms align perfectly with our derived strict lower bounds.





{\footnotesize \bibliographystyle{acm}
\bibliography{references}}

\appendix

\section{Proof regarding Linear Estimator}
\subsection{Proof of Theorem \ref{thm:var_for_any_P}}
\label{sec:proof_thm:var_for_any_P}
The output of each object is represented using one-hot encoding. For an original value $x$, let $Y_x = \sum_{i \in [n]} M(X^{(i)}) \mathbf{1}(X^{(i)} = x)$, where $\mathbf{1}(\cdot)$ is the indicator function. The $o$-th element of $Y_x$ represents the count of responses equal to $o$ generated by the objects with value $x$. Since there are $nf(x)$ objects in the dataset with value $x$, $Y_x$ is a multinomial random variable distributed as $\text{MN}(n f(x), P_{*, x})$, where $P_{*, x}$ denotes the $x$-th column of the perturbation matrix $P$. Consequently, $n \mathbf{f_O} = \sum_{x \in [d]} Y_x$, and its covariance is given by:
\begin{multline*}
    \text{Cov}(\mathbf{f_O}) = \sum_{x \in [d]} f(x) (\text{diag}(P_{*, x}) - P_{*, x} P_{*, x}^T )  \\
    = \text{diag}(P \mathbf{f}) - P \text{diag}(\mathbf{f}) P^T ,
\end{multline*}
\noindent where $\mathbf{f} = [f(1), f(2), \dots, f(d)]^T$. Given that the covariance of a linear transformation is $\text{Cov}(Q\mathbf{f_O}) = Q \text{Cov}(\mathbf{f_O}) Q^T$, we obtain:
\begin{multline*}
    \text{Cov}(Q \mathbf{f_O}) = Q \text{diag}(P \mathbf{f}) Q^T - Q P \text{diag}(\mathbf{f}) P^T Q^T \\
    =  Q \text{diag}(P \mathbf{f}) Q^T -   \text{diag}(\mathbf{f}) .
\end{multline*}

\noindent Since $\text{Tr}(Q \text{diag}(P \mathbf{f}) Q^T ) = (Q \circ Q) P \mathbf{f}$, it follows that:
\begin{equation*}
    \text{Var}(Q \mathbf{f_O}) = \text{Tr}(\text{Cov}(Q \mathbf{y})) = (Q \circ Q) P  \mathbf{f} - \mathbf{f} . \qed
\end{equation*}

\subsection{Proof of Theorem \ref{thm:var_theta}}
\label{sec:proof_thm:var_theta}
Every $M(X^{(i)})$ is sampled i.i.d. from a category distribution with probability $P \bm{\theta}$, which derives that
\begin{equation*}
    \text{Cov}(\sum M(X^{(i)})) = n [ \text{diag}(P \bm{\theta}) - P \bm{\theta} \bm{\theta}^TP^T  ] .
\end{equation*}
\noindent Following Appendix \ref{sec:proof_thm:var_for_any_P}, we prove that
\begin{equation*}
    \text{Var}(Q \mathbf{f_O})  = (Q \circ Q) P  \bm{\theta} - \bm{\theta}^2 . \qed
\end{equation*}

\subsection{Proof of Theorem \ref{thm:l1_approx} and Corollary \ref{cor:l1_to_l2}} 
\label{sec:proof_thm:l1_approx}
This proof focuses on frequency estimation, but it is applicable to distribution estimation. Reusing the notation from Appendix \ref{sec:proof_thm:var_for_any_P}, the output frequency is expressed as $\mathbf{f_O} = \frac{1}{n} \sum_{x \in [d]} Y_x = \frac{1}{n} \sum_{x \in [d]} \text{MN}(n f(x), P_{*, x})$, where $\text{MN}$ denotes a multinomial distribution and $P$ is the perturbation matrix. Given that $\hat{f}(x) = Q_{x, *} \mathbf{f_O}$, where $Q_{x, *}$ is the $x$-th row of $Q$, we have:
\begin{equation*}
    \hat{f}(x) = \frac{1}{n} Q_{x, *} \sum_{x \in [d]} \text{MN}(n f(x), P_{*, x}) .
\end{equation*}

For sufficiently large $n$, the Central Limit Theorem (CLT) ensures that the linear combination of these independent multinomial random variables can be approximated by a Normal distribution with the same expectation and variance. Since the Mean Absolute Error (MAE) of a Normal distribution is $\sqrt{\frac{2}{\pi} \sigma^2}$, where $\sigma^2$ is its variance, it follows that:
\begin{equation*}
    \lim_{n \rightarrow \infty}\mathbb{E}[| \hat{f}(x) - f(x) |] = \sqrt{\frac{2}{\pi} \text{Var}(\hat{f}(x))} .
\end{equation*}

\noindent Consequently,
\begin{equation*}
    \lim_{n \rightarrow \infty} \sum_{x \in [d]}  \mathbb{E}[| \hat{f}(x) - f(x) |] = \sum_{x \in [d]} \sqrt{\frac{2}{\pi} \text{Var}(\hat{f}(x))} ,
\end{equation*}
\noindent which proves Theorem \ref{thm:l1_approx} $\qed$

To prove Corollary \ref{cor:l1_to_l2}, we apply Theorem \ref{thm:perm_var_aI+bJ}, yielding $\text{Var}(\hat{f}(x)) = (\alpha - \beta - 1)f(x) + \beta$. Note that $\sum_x \sqrt{\text{Var}(\hat{f}(x))}$ is a concave function of $\mathbf{f}$. Given the constraint $\sum_x f(x) = 1$, Jensen's inequality guarantees that $\sum_x \sqrt{\text{Var}(\hat{f}(x))}$ is maximized when $f(x) = \frac{1}{d}$ for all $x \in [d]$. Therefore,
\begin{multline*}
    \lim_{n \rightarrow \infty} \mathcal{L}_1(\hat{f}) = \lim_{n \rightarrow \infty}   \max_{\forall \mathbf{X}: |\mathbf{X}| = n} \sum_{x \in [d]}  \mathbb{E}[| \hat{f}(x) - f(x) |] \\ = d\sqrt{\frac{2}{\pi} \text{Var}(\hat{f}(x) | f(x) = \frac{1}{d}) } \\ = d \sqrt{ \frac{2}{\pi} [(\alpha - \beta - 1) \frac{1}{d} + \beta]} =  \sqrt{ \frac{2 d}{\pi} \mathcal{L}_2}  . \qed
\end{multline*}

\section{Proof of Precision Optimality}

\subsection{Proof of Theorem \ref{thm:L2_fisher}}
\label{sec:proof_thm:L2_fisher}
Let $\bm{\theta} \in \mathbb{R}^d$ denote the parameter vector of a categorical distribution $\mathcal{C}(\bm{\theta})$. Define the random variable $O^{(i)}$ as the $i$-th output of $M(\mathcal{C}(\bm{\theta}))$, where mechanism $M$ has a perturbation matrix $P$. The probability of observing output value $o$ is given by $\Pr(O^{(i)} = o) = \mathbf{e}_{o}^T P \bm{\theta}$, where $\mathbf{e}_{o}$ is the $o$-th unit basis vector. The score function $D(O^{(i)})$ is defined as:
\begin{align*}
D(O^{(i)}) &= \nabla_{\bm{\theta}} \log (\Pr(O^{(i)})) = \frac{\nabla_{\bm{\theta}} (\mathbf{e}_{O^{(i)}}^T P  \bm{\theta})}{\Pr(O^{(i)})} \\
&= \frac{P^T \mathbf{e}_{O^{(i)}}}{\Pr(O^{(i)})} = P^T \text{diag}(P \bm{\theta})^{-1} \mathbf{e}_{O^{(i)}}.
\end{align*}

\noindent  The expectation of the score function is:
\begin{multline}
    \mathbb{E}[D(O^{(i)})] = \sum_{o \in [m]} \Pr(o) \frac{P^T \mathbf{e}_{o}}{\Pr(o)} \\ = P^T \sum_{o \in [m]} \mathbf{e}_{o} = P^T \mathbf{1}_m = \mathbf{1}_d.
    \label{eq:exp_score}
\end{multline}
\noindent The second moment of the unit vectors is given by:
\begin{equation*}
    \mathbb{E} [\mathbf{e}_{O^{(i)}} \mathbf{e}_{O^{(i)}}^T] = \sum_{o \in [m]} \Pr(o) \mathbf{e}_o \mathbf{e}_o^T = \text{diag}(P \bm{\theta}),
\end{equation*}
\noindent which yields the second moment of the score function:
\begin{multline}
    \mathbb{E}[D(O^{(i)}) D(O^{(i)})^T]  \\ = P^T \text{diag}(P \bm{\theta})^{-1} \mathbb{E} [\mathbf{e}_{O^{(i)}} \mathbf{e}_{O^{(i)}}^T] \text{diag}(P \bm{\theta})^{-1} P  \\
    = P^T \text{diag}(P \bm{\theta})^{-1} P.
    \label{eq:second_moment}
\end{multline}

Since $O^{(i)}$ and $O^{(j)}$ are independent for $i \ne j$, their centered score functions are uncorrelated:
\begin{equation}
    \mathbb{E}[(D(O^{(i)}) - \mathbf{1}_d) (D(O^{(j)}) - \mathbf{1}_d)^T] = 0.
    \label{eq:uncorrelated}
\end{equation}

Define the aggregate centered score function as $\overline{D}(\mathbf{O}) = \sum_{i=1}^n [D(O^{(i)}) - \mathbf{1}_d]$. From Eqs. \eqref{eq:exp_score}, \eqref{eq:second_moment}, and \eqref{eq:uncorrelated}, it follows that $\mathbb{E}[\overline{D}(\mathbf{O})] = \mathbf{0}$ and its covariance matrix is:
\begin{multline}
    \mathbb{E}[\overline{D}(\mathbf{O}) \overline{D}(\mathbf{O})^T] = n (P^T \text{diag}(P \bm{\theta})^{-1} P - \mathbf{1}_d \mathbf{1}_d^T) \\ \overset{\text{def}}{=} n( \mathscr{I}_U - \mathbf{1}_d \mathbf{1}_d^T ) \overset{\text{def}}{=}  n \mathscr{I}_C , \label{eq:E_DO_DO}
\end{multline}
\noindent where $\mathscr{I}_U$ and $\mathscr{I}_C$ refer to the unconstrained and constrained Fisher Information Matrices (FIM), respectively.

We define $\mathscr{I}_C^{+} = \mathscr{I}_U^{-1} - \bm{\theta} \bm{\theta}^T $ as the generalized inverse of the constrained FIM. Note that:
\begin{align}
    & \mathscr{I}_U \bm{\theta} = P^T \text{diag}(P \bm{\theta})^{-1} P \bm{\theta} = P^T \mathbf{1}_m = \mathbf{1}_d , \notag \\
    & \mathscr{I}_U^{-1} \mathbf{1}_d = \bm{\theta} . \label{eq:Ic_theta_one}
\end{align}
\noindent Using them with $\bm{\theta}^T \mathbf{1}_d = 1$, we derive:
\begin{multline*}
    \mathscr{I}_C^{+} \mathscr{I}_C = \mathscr{I}_U^{-1} \mathscr{I}_U - \mathscr{I}_U^{-1} \mathbf{1}_d \mathbf{1}_d^T - \bm{\theta} \bm{\theta}^T \mathscr{I}_U  + \bm{\theta} \bm{\theta}^T \mathbf{1}_d \mathbf{1}_d^T \\ = I_d - \bm{\theta} \mathbf{1}_d - \bm{\theta} \mathbf{1}_d + \bm{\theta} \mathbf{1}_d^T =  I_d - \bm{\theta} \mathbf{1}_d  , 
\end{multline*}
\noindent which leads to:
\begin{multline}
    \mathscr{I}_C^{+} \mathscr{I}_C \mathscr{I}_C^{+} = \mathscr{I}_U^{-1} - \bm{\theta} \bm{\theta}^T - \bm{\theta} \mathbf{1}_d^T \mathscr{I}_U^{-1} + \bm{\theta} \mathbf{1}_d  \bm{\theta} \bm{\theta}^T \\ = \mathscr{I}_U^{-1} - \bm{\theta} \bm{\theta}^T - \bm{\theta}  \bm{\theta} ^T + \bm{\theta}  \bm{\theta} ^T  = \mathscr{I}_C^{+}. \label{eq:IcIcIc}
\end{multline}

Let $\hat{\bm{\theta}}$ be the estimated distribution of $\bm{\theta}$ from an unbiased estimator $\hat{\theta}$. We define the cross-covariance matrix $C$ as:
\begin{align}
    C &= \mathbb{E} [ (\hat{\bm{\theta}} - \bm{\theta}) \overline{D}(\mathbf{O})^T ] = \mathbb{E} [ \hat{\bm{\theta}} \overline{D}(\mathbf{O})^T ] \label{eq:C_def}\\
    &= \mathbb{E} \left[ \hat{\bm{\theta}} \sum_{i \in [n]} ( \nabla_{\bm{\theta}} \log \Pr(O^{(i)})^T - \mathbf{1}^T ) \right]  \notag \\
    &= \sum_{\mathbf{o} \in [m]^n} \Pr(\mathbf{o}) \hat{\bm{\theta}} \frac{ \nabla_{\bm{\theta}} \Pr(\mathbf{o}) }{\Pr(\mathbf{o})} - n \mathbb{E} [ \hat{\bm{\theta}} \mathbf{1}^T ] \notag \\
    &= \nabla_{\bm{\theta}} \sum_{\mathbf{o} \in [m]^n} \hat{\bm{\theta}} \Pr(\mathbf{o}) - n \bm{\theta} \mathbf{1}^T = \nabla_{\bm{\theta}} \mathbb{E} [ \hat{\bm{\theta}} ] - n \bm{\theta} \mathbf{1}^T = I_d - n \bm{\theta} \mathbf{1}^T. \notag 
\end{align}
Using Eqs. \eqref{eq:Ic_theta_one} and $\bm{\theta}^T \mathbf{1}_d = 1$, we have:
\begin{multline}
    C \mathscr{I}_C^{+} = \mathscr{I}_U^{-1} - n \bm{\theta} \mathbf{1}^T \mathscr{I}_U^{-1} - \bm{\theta}  \bm{\theta} ^T + n \bm{\theta} \mathbf{1}^T \bm{\theta} \bm{\theta}^T \\ = \mathscr{I}_U^{-1} - n \bm{\theta} \bm{\theta}^T - \bm{\theta}  \bm{\theta} ^T + n \bm{\theta} \bm{\theta}^T = \mathscr{I}_C^{+}  .
    \label{eq:C_I_c=I_c}
\end{multline}

Let $V = \mathbb{E}[ (\hat{\bm{\theta} } - \bm{\theta}) (\hat{\bm{\theta}} - \bm{\theta})^T ]$ denote the covariance matrix of $\hat{\bm{\theta}}$. We consider an auxiliary vector $\mathbf{g} = (\hat{\bm{\theta}} - \bm{\theta}) - \frac{1}{n}  \mathscr{I}_C^{+} \overline{D}(\mathbf{O})$. Since the expectation of the outer product of any random vector is positive semi-definite, and given Eqs. \eqref{eq:E_DO_DO} and \eqref{eq:C_def},  it follows that:
\begin{equation*}
    \mathbb{E}[\mathbf{g} \mathbf{g}^T] = V - \frac{1}{n} C \mathscr{I}_C^{+}  - \frac{1}{n}  \mathscr{I}_C^{+} C^T + \frac{1}{n}  \mathscr{I}_C^{+} \mathscr{I}_C \mathscr{I}_C^{+}  \succeq 0.
\end{equation*}

Using $C \mathscr{I}_C^{+} = \mathscr{I}_C^{+}$ and $\mathscr{I}_C^{+} \mathscr{I}_C \mathscr{I}_C^{+} = \mathscr{I}_C^{+}$ from Eqs. \eqref{eq:IcIcIc} and \eqref{eq:C_I_c=I_c}, this simplifies to $V \succeq \frac{1}{n} \mathscr{I}_C^{+}$. Therefore:
\begin{align*}
    L_2(\hat{\theta}, \bm{\theta}, n) &= \text{Tr}(V) \ge \frac{1}{n} \text{Tr}(\mathscr{I}_C^{+}) \\
    &= \frac{1}{n} \text{Tr} \left[ (P^T \text{diag}(P \bm{\theta})^{-1} P)^{-1} - \bm{\theta} \bm{\theta}^T \right]. \qed
\end{align*}

For the $L_1$ lower bound, when the dataset size $n$ is asymptotically large, the Hájek-Le Cam Asymptotic Minimax Theorem states that the bowl-shaped loss, e.g., $L_1$, of any estimator is lower bounded by that of a normal distribution with covariance $\frac{1}{n}\mathscr{I}_C^{+}$, the inverse FIM. For a normal distribution, $\mathbb{E}[ | \hat{\theta}_x - \theta_x | ] = \sqrt{\frac{2}{n \pi} \text{Var}(\hat{\theta}_x)}$. Thus, the $L_1$ loss is bounded by:
\begin{equation*}
    \lim_{n \rightarrow \infty} L_1(\hat{\theta}, \bm{\theta}, n) \ge \sum_{x \in [d]} \sqrt{\frac{2}{n \pi} (\mathscr{I}_C^{+})_{x,x}} . \qed
\end{equation*}

\subsection{Proof of Theorem \ref{thm:var_theta_to_f}}
\label{sec:proof_thm:var_theta_to_f}
For any $x \in [d]$, denote $\hat{\theta} = \hat{f}$ as the $x$-th estimated value of $\hat{\bm{\theta}} = \hat{\mathbf{f}}$, and $f$ and $\theta$ are the $x$-th elements of the sampled frequency $\mathbf{f}$ and the distribution $\bm{\theta}$. We have
\begin{multline}
    \text{Var}(\hat{\theta}) = \mathbb{E}[ (\hat{f} - f + f - \theta)^2 ] \\ =  \mathbb{E}[ (\hat{f} - f)^2 ]  + \mathbb{E}[ (f - \theta)^2 ]  + 2 \mathbb{E}[ (\hat{f} - f) (f - \theta) ]. \label{eq:var_f_theta_step_1}
\end{multline}

Subsequently, we derive
\begin{multline*}
    \mathbb{E}[ (\hat{f} - f)^2 ] = \sum_{\mathbf{f}}  \Pr(\mathbf{f} | \bm{\theta}) \sum_{\hat{f}} \Pr(\hat{f} | \mathbf{f})  (\hat{f} - f)^2 
    \\ =  \sum_{\mathbf{f}}  \Pr(\mathbf{f} | \bm{\theta}) \text{Var}(\hat{f} | \mathbf{f}) .
\end{multline*}

Given the category distribution, $\mathbb{E}[ (f - \theta)^2 ] = \frac{\theta - \theta^2}{n}$. Additionally, given that $\mathbb{E}[ \hat{f} - f ] = 0$
\begin{multline*}
    \mathbb{E}[ (\hat{f} - f) (f - \theta) ] = \sum_{\mathbf{f}}  \Pr(\mathbf{f} | \bm{\theta})\sum_{\hat{f}} \Pr(\hat{f} | \mathbf{f})  (\hat{f} - f) (f - \theta) \\
    = \sum_{\mathbf{f}}  \Pr(\mathbf{f} | \bm{\theta}) \mathbb{E}[\hat{f} - f | f] (f - \theta) = 0 .
\end{multline*}

Substituting them back to Eq. \eqref{eq:var_f_theta_step_1}, we obtain 
\begin{equation*}
     \text{Var}(\hat{\theta}) = \sum_{\mathbf{f}}  \Pr(\mathbf{f} | \bm{\theta}) \text{Var}(\hat{f} | \mathbf{f}) +  \frac{\theta - \theta^2}{n} . \qed
\end{equation*}

\subsection{Proof of Theorem \ref{thm:constant_support_optimality}}
\label{sec:proof_thm:constant_support_optimality}
Recalling Eq. \eqref{eq:extremal_matrix_form}, we have
\begin{multline*}
    V = P^T  \overline{P}^{-1} P^T  = [(e^{\epsilon} - 1) S^T + J^T ] \frac{P_O^2}{\overline{P}} [(e^{\epsilon} - 1) S + J ] \\
    = [(e^{\epsilon} - 1) S^T + J^T ] \frac{P_O}{\frac{1}{d} \text{diag}((e^{\epsilon} - 1)\sum_x S_{o, x} + d)} [(e^{\epsilon} - 1) S + J ] \\
    \overset{\mathrm{def}}{=} [(e^{\epsilon} - 1) S^T + J^T ] \frac{P_O}{B} [(e^{\epsilon} - 1) S + J ] 
\end{multline*}
\noindent where $B = \frac{1}{d} \text{diag}((e^{\epsilon} - 1)\sum_x S_{o, x} +1)$.

The trace of $V^{-1}$ is the sum of its eigenvalues, each of which is the reciprocal of an eigenvalue of $V$. Thus, we investigate the eigenvalues of $V$. Notably, $[(e^{\epsilon} - 1) S + J ] \mathbf{1} = d [b_1, b_2, \dots, b_m]^T$, where $\mathbf{1}$ represents the all-ones vector. Since $[(e^{\epsilon} - 1) S^T + J^T ] P_O \mathbf{1} = P^T \mathbf{1} = \mathbf{1}$, we have:
\begin{multline*}
    P^T  \overline{P}^{-1} P \mathbf{1} = P^T B^{-1} [(e^{\epsilon} - 1) S + J ]  \mathbf{1} \\= P^T B^{-1} d [b_1, b_2, ..., b_m]^T = d P^T  \mathbf{1}  = d \mathbf{1} .
\end{multline*}
\noindent Consequently, $\mathbf{1}$ is an eigenvector of $V$ with an associated eigenvalue of $d$, which we refer to as the first eigenvalue. It follows that $\text{Tr}(V) = d + \sum_{i = 2}^d \lambda_i$, where $\lambda_i$ represents the $i$-th eigenvalue. Similarly, $\text{Tr}(V^{-1}) = \frac{1}{d} + \sum_{i=2}^d \frac{1}{\lambda_i}$. For a fixed $\text{Tr}(V)$, $\text{Tr}(V^{-1})$ is minimized when the remaining eigenvalues $[\lambda_i]_{i \ge 2}$ are equal to one another. \qed

We now expand $\text{Tr}(V)$:
\begin{multline*}
    \text{Tr}(S^T \frac{P^O}{ K } S ) = \sum_x \sum_o \frac{p_o}{ \frac{1}{d} [(e^{\epsilon} - 1)\sum_x S_{o, x} + d] }  S_{o, x}^2 \\ =  \sum_o \frac{p_o \sum_x  S_{o, x} }{ \frac{1}{d} (e^{\epsilon} - 1)\sum_x S_{o, x} + 1 }  
\end{multline*}
\begin{multline*}
    \text{Tr}(S^T \frac{P^O}{ B } J ) = \sum_x \sum_o \frac{p_o}{ \frac{1}{d} [(e^{\epsilon} - 1)\sum_x S_{o, x} + d] }  S_{o, x} \\ =  \sum_o \frac{p_o \sum_x  S_{o, x} }{ \frac{1}{d} (e^{\epsilon} - 1)\sum_x S_{o, x} + 1 } 
\end{multline*}
\begin{multline*}
    \text{Tr}(J^T \frac{P^O}{ B } J ) =  \sum_x \sum_o \frac{p_o}{ \frac{1}{d} [(e^{\epsilon} - 1)\sum_x S_{o, x} + d] }  d \\ =  \sum_o \frac{p_o   d}{ \frac{1}{d} (e^{\epsilon} - 1)\sum_x S_{o, x} + 1 } 
\end{multline*}

Substituting these terms into $\text{Tr}(V)$, we obtain:
\begin{equation*}
    \text{Tr}(V) = d \sum_o p_o \frac{(e^{2\epsilon} - 1) \sum_x S_{o, x} + d}{(e^{\epsilon} - 1) \sum_x S_{o, x} + d} .
\end{equation*}
\noindent Let $k_o = \sum_x S_{o, x}$. It can be verified that $\text{Tr}(V)$ is a concave function with respect to each $k_o$. Furthermore, the extremal configuration requires $\sum_x \sum_o p_o S_{o, x} = \sum_o p_o k_o = \frac{d p^*}{e^{\epsilon}}$ (by Lemma \ref{lem:constant_sp}). The maximum of $\text{Tr}(V)$ is achieved when $k_o$ is constant for all $o$, i.e., $\forall o: k_o = \frac{d p^*}{e^{\epsilon} \sum_o p_o} = k$. Consequently, we have:
\begin{multline*}
    V = \frac{1}{(e^{\epsilon} - 1) k + 1} [(e^{\epsilon} - 1) S^T + J^T ] P_O [(e^{\epsilon} - 1) S + J ] \\ =  \frac{1}{(e^{\epsilon} - 1) k + 1} [(e^{\epsilon} - 1) S^T + J^T ] P.
\end{multline*}

Notably, since $S^T P = (p^* - q^*) I + q^* J$ and $J^T P = J$, $V$ also takes the form $a I + bJ$, where $a$ and $b$ are constants. Aside from the first eigenvalue being fixed at $d$, all other eigenvalues are equal in this form. Thus, $\text{Tr}(V^{-1})$ also reaches its minimum. $\qed$

\subsection{Proof of Theorem \ref{thm:optimized_urp_reconstruction}}
\label{sec:proof_thm:optimized_urp_reconstruction}
Assuming a constant support size, i.e., $\forall o: \sum_x S_{o, x} = k$, we have $\overline{P}^{-1} = \frac{1}{(e^{\epsilon} - 1)k + d} P_O^{-1}$. The optimized reconstruction matrix $Q^*$ is expressed as
\begin{multline*}
    Q^* = (P^T \overline{P}^{-1} P)^{-1} P^T \overline{P}^{-1} 
    \\ = ([(e^{\epsilon} - 1) S^T + J] P)^{-1} [(e^{\epsilon} - 1) S^T + J] .
\end{multline*}
\noindent For each permutation $i \in [D]$, we have:
\begin{align*}
    Q^*_i &= Z_i^T (P^T \overline{P}^{-1} P)^{-1} P^T \overline{P}^{-1} \\
    &= Z_i^T(P^T \overline{P}^{-1} P)^{-1} Z_i Z_i^T P^T \overline{P}^{-1} \\
    &= (Z_i^T P^T \overline{P}^{-1} P Z_i)^{-1} Z_i^T [(e^{\epsilon} - 1) S^T + J] \\
    &= (Z_i^T P^T \overline{P}^{-1} P Z_i)^{-1} [(e^{\epsilon} - 1) (S Z_i)^T + J].
\end{align*}

\noindent As proven in Appendix \ref{sec:proof_thm:constant_support_optimality}, $P^T \overline{P}^{-1} P$ takes the form $a I + bJ$. Consequently, $Z_i^T P^T \overline{P}^{-1} P Z_i = P^T \overline{P}^{-1} P$, which implies:
\begin{equation*}
    Q^*_i = (P^T \overline{P}^{-1} P)^{-1}  [(e^{\epsilon} - 1) (S Z_i)^T + J].
\end{equation*}
Furthermore, let $Q_{U}$ be defined as:
\begin{equation*}
    Q_{U} = (P^T \overline{P}^{-1} P)^{-1} [(e^{\epsilon} - 1) S_{U}^T + J] ,
\end{equation*}
\noindent where $S_{U} = [S_1, S_2, \dots, S_D]^T$. Since $Q_{U} P_{U} = I$, it follows directly that $P_{U}^T \overline{P}_{U}^{-1} P_{U} = P^T \overline{P}^{-1} P$. Thus:
\begin{multline*}
    Q_{U} = (P_{U}^T \overline{P}_{U}^{-1} P_{U})^{-1} [(e^{\epsilon} - 1) S_{U}^T + J] \\
    = [\left(  (e^{\epsilon} - 1) S_U^T + J \right) P_U]^{-1} [(e^{\epsilon} - 1) S_{U}^T + J] 
\end{multline*}

In the remainder of this section, we abbreviate $S_{U}$ and $P_{U}$ as $S$ and $P$, respectively.

First, consider the term $(e^{\epsilon} - 1) S^T + J$. Eq. \eqref{eq:S^T_M_matrix} demonstrates that $S^T \sum_i U(X^{(i)}) = [\sum_{i \in [n]} S_{O^{(i)}, x}]_{x \in [d]}$. Thus, $(e^{\epsilon} - 1) S^T + J$ computes $y_x = (e^{\epsilon} - 1) \frac{\sum_i S_{O^{(i)}, x}}{n} + 1$ for every $x$.

Next, we analyze $[((e^{\epsilon} - 1) S^T + J) P]^{-1}$. Since URP guarantees a symmetric configuration (Theorem \ref{thm:URP_sym}), we have $S^T P = (p^* - q^*) I + q^* J$. Given that $J P = J$, it follows that:
\begin{equation*}
    \left( (e^{\epsilon} - 1) S^T + J \right) P = (e^{\epsilon} - 1) (p^* - q^*) I + \left( (e^{\epsilon} - 1) q^* + 1 \right) J .
\end{equation*}

Applying the Sherman-Morrison formula, we obtain:
\begin{multline*}
    [\left( (e^{\epsilon} - 1) S^T + J \right) P]^{-1} = \frac{1}{(e^{\epsilon} - 1) (p^* - q^*)} I \\ + \frac{q(e^{\epsilon} - 1) + 1}{(e^{\epsilon} - 1) (p^* - q^*) \sum_x y_x} J,
\end{multline*}
\noindent where $\sum_x y_x = d[(e^{\epsilon} - 1) q^* + 1] + (e^{\epsilon} - 1) (p^* - q^*)$. Note that $\sum_x y_x$ is constant because $Q^*$ requires a constant support size, i.e., each response supports the same number of original values, and $\sum_x y_x = \mathbf{1}^T [(e^{\epsilon} - 1) S^T + J] P$. 

Summarizing these steps:
\begin{enumerate}
    \item Compute $y_x = (e^{\epsilon} - 1) \frac{\sum_i S_{O^{(i)}, x}}{n} + 1$ for each element.
    \item Estimate the frequency as:
    \begin{equation*}
        \hat{f}(x) = \frac{y_x}{(e^{\epsilon} - 1)(p^* - q^*)} - \frac{q^*(e^{\epsilon} - 1) + 1}{(e^{\epsilon} - 1) (p^* - q^*) \sum_x y_x} \sum_x y_x.
    \end{equation*}
\end{enumerate}
\noindent This yields $\hat{f}(x) = \frac{(\sum_{i \in [n]} S_{o^{(i)}, x}) / n - q^* }{p^* - q^*}$, identical to Eq. \eqref{eq:sym_fx}. $\qed$

\subsection{The proof of Lemma \ref{lem:p_q_star_k}}
\label{sec:proof_lem:p_q_star_k}
Let $d$, $\epsilon$, and $n$ denote the dictionary size, the privacy parameter, and the dataset size, respectively. Suppose we have a LDP estimator that satisfies the optimal symmetric configuration with support size $k$. Let there be $m$ possible output responses from its perturbation, denoted by $o \in [m]$. We define $p_o$ as the base probability of output $o$, and $S_{o, x} \in \{0, 1\}$ as the indicator that $o$ supports $x \in [d]$. For a symmetric configuration, the parameters $p^*$ and $q^*$ can be expressed as:
\begin{equation}
   \forall x \in [d]:  p^* = \sum_o^m S_{o, x} e^{\epsilon} p_o \label{eq:p_star}
\end{equation}
\begin{multline}
   \forall x_1 \ne x_2 \in [d], q^* = \sum_o^m S_{o, x_1} S_{o, x_2}  e^{\epsilon} p_o + (1 - S_{o, x_1}) S_{o, x_2} p_o \\
   = (e^{\epsilon} - 1) \sum_o^m S_{o, x_1} S_{o, x_2}  p_o + \sum_o^m  S_{o, x_2} p_o \label{eq:q_star}
\end{multline}

First, we derive the expression for $p^*$ in Lemma \ref{lem:p_q_star_k} using the following lemma:
\begin{lemma}
    \begin{align*}
        \forall x \in [d]: &  \sum_o^m S_{o, x}  p_o  =  \frac{ k }{ k (e^{\epsilon} - 1) + d} ,\\
        & \sum_o^m p_o = \frac{d}{ k (e^{\epsilon} - 1) + d} .
    \end{align*}
\label{lem:sp_and_p}
\end{lemma}

To prove this lemma, consider the following matrix:
\begin{equation*}
    \begin{pmatrix}
S_{1, 1} p_1 & S_{1, 2} p_1 & ... & S_{1, d} p_1\\
S_{2, 1} p_2 & ... & ... & S_{2, d} p_2 \\
...
\\S_{m, 1} p_m  & ... & ... & S_{m, d} p_m
\end{pmatrix} 
\end{equation*}

The sum of all elements in the matrix can be calculated by either summing over rows or summing over columns. Since every response has a constant support size $k$, we have $\forall o: \sum_{x=1}^d S_{o, x} = k$. Thus:
\begin{equation}
    d  \sum\limits_{o}^m S_{o, x}  p_o =  k \sum\limits_{o}^m  p_o \label{eq:sp_and_p_eq_1}
\end{equation}
Furthermore, for any input $x$, the total probability of all possible outputs must sum to one:
\begin{equation}
    \forall x \in [d]: (e^{\epsilon} - 1)  \sum_{o}^m S_{o, x}  p_o +  \sum\limits_{o}^m  p_o = 1 . \label{eq:sum_p_to_one}
\end{equation}

Given that the extremal configuration requires $\sum_{o=1}^m S_{o, x} p_o$ to be constant for all $x$, we can solve Eqs. \eqref{eq:sp_and_p_eq_1} and \eqref{eq:sum_p_to_one} simultaneously to obtain $\sum_{o=1}^m S_{o, x} p_o$ and $\sum_{o=1}^m p_o$ stated in Lemma \ref{lem:sp_and_p}. $\qed$

Lemma \ref{lem:sp_and_p} together with Eq. \eqref{eq:p_star} yields the formula for $p^*$ in Lemma \ref{lem:p_q_star_k}. We now derive $q^*$. Since $\sum_{o=1}^m S_{o, x} p_o$ is a known constant, solving for $q^*$ is equivalent to determining $\sum_{o=1}^m S_{o, x_1} S_{o, x_2} p_o$. Consider the support matrix:
\begin{equation*}
    \begin{pmatrix}
S_{1, 1} & S_{1, 2} & ... & S_{1, d} \\
S_{2, 1} & ... & ... & S_{2, d} \\
...
\\S_{m, 1} & ... & ... & S_{m, d} 
\end{pmatrix} .
\end{equation*}

Suppose we sample a row $o$ from $S$ with probability $p_o / \sum_{j=1}^m p_j$. Let $Y_i$ be a random variable representing the $i$-th element of the sampled row. Based on Lemma \ref{lem:sp_and_p}, we have:
\begin{equation*}
    \text{Pr}( Y_i = 1 ) = \frac{ \sum_{o}^m S_{o, i} p_o  }{ \sum_{o}^m p_o  } = \frac{k}{d} .
\end{equation*}

\noindent Its variance and covariances are given by:
\begin{equation*}
    \text{Var}( Y_i ) = \text{Pr}( Y_i = 1 ) (1 - \text{Pr}( Y_i = 1 )) = \frac{ k }{d} (1 - \frac{k}{d}) .
\end{equation*}
\begin{equation*}
\begin{aligned}
    \text{Cov}(Y_{i_1}, Y_{i_2}) &= \text{Pr}(Y_{i_1} = 1, Y_{i_2} = 1) - \text{Pr}(Y_{i_1} = 1)\text{Pr}(Y_{i_2} = 1) \\
    &= \frac{\sum_{o=1}^m S_{o, i_1} S_{o, i_2} p_o}{\sum_{o=1}^m p_o} - \left(\frac{k}{d}\right)^2 \overset{\text{def}}{=} \overline{c} - \left(\frac{k}{d}\right)^2.
\end{aligned}
\end{equation*}
Let $V_{\Sigma} = \text{Var}(\sum_{i=1}^d Y_i)$. Since $\sum_i Y_i$ represents the support size $k$, which is constant, $V_{\Sigma} = 0$. Expanding the variance of the sum:
\begin{equation*}
\begin{aligned}
    V_{\Sigma} &= \sum_{i=1}^d \text{Var}(Y_i) + \sum_{i_1=1}^d \sum_{i_2 \neq i_1} \text{Cov}(Y_{i_1}, Y_{i_2}) \\
    &= d \cdot \frac{k}{d} \left(1 - \frac{k}{d}\right) + d(d-1) \left(\overline{c} - \left(\frac{k}{d}\right)^2\right) = 0.
\end{aligned}
\end{equation*}

\noindent Solving for $\overline{c}$, we find:

\begin{equation}
    \overline{c} = \frac{ k ( k - 1) }{d (d - 1)} . \label{eq:lb_c}
\end{equation}
By applying Lemma \ref{lem:sp_and_p} to $\overline{c}$ and substituting the result into Eq. \eqref{eq:q_star}, we derive $q^*$ in Lemma \ref{lem:p_q_star_k}. $\qed$

\subsection{Proof of Theorem \ref{thm:optimized_l1_l2}}
\label{sec:proof_thm:optimized_l1_l2}
Substituting Eq. \eqref{eq:var_fx_as_k} into $\mathcal{L}_2(\hat{f}, n)$, we obtain
\begin{multline}
    \mathcal{L}_2(\hat{f}, n) = \max_{\mathbf{f}} d \frac{ k (e^{\epsilon} - 1) + d -1 ] [k (e^{\epsilon} - 1) + d - e^{\epsilon} ]  }{ n k (d - k)  (e^{\epsilon} - 1)^2}  \\
        +  \frac{k^2(e^{\epsilon} - 1) + k (2d - e^{\epsilon} - 1) + d - d^2 }{ n  k (d - k)  (e^{\epsilon} - 1)} \label{eq:l2_k}
\end{multline}

Since the parameters are independent of any $\mathbf{f}$, we can remove the maximum operator. Differentiating Eq. \eqref{eq:l2_k} with respect to $k$, we acquire
\begin{equation*}
    \frac{\partial \mathcal{L}_2(\hat{f}, n)}{\partial k} = \frac{(d - 1)^2 (k e^{\epsilon} - k + d) (k e^{\epsilon} + k - d)}{n k^2 (k - d)^2 (e^{\epsilon} - 1)^2} .
\end{equation*}

\noindent Considering $k \ge 1$, we find $\frac{\partial \mathcal{L}_2(\hat{f})}{\partial k} = 0$ when $k = \frac{d}{e^{\epsilon} + 1}$. The optimality can be confirmed by verifying that $\frac{\partial^2 \mathcal{L}_2(\hat{f})}{\partial k^2} > 0$ at $k = \frac{d}{e^{\epsilon} + 1}$. Substituting $k = \frac{d}{e^{\epsilon} + 1}$ back into Eq. \eqref{eq:l2_k} yields the expression for $\mathcal{L}_2^*$ in Theorem \ref{thm:optimized_l1_l2}. The bound $\mathcal{L}_1^*$ is derived following Corollary \ref{cor:l1_to_l2}. Thus, Theorem \ref{thm:optimized_l1_l2} is proven. $\qed$

\subsection{Proof of Corollary \ref{cor:l1_l2_deviation}}
\label{sec:proof_cor:l1_l2_deviation}
Recall that $\mathcal{L}_2$ in Eq. \eqref{eq:l2_k} is a function of $k$; hence, we denote it as $\mathcal{L}_2(k)$. If the optimal value $k^* = \frac{d}{e^{\epsilon} + 1}$ is not an integer, it deviates by at most $\frac{1}{2}$ from the nearest integers. We define the deviation function as
\begin{equation}
    \Delta \mathcal{L}_2 (\Delta k )= \frac{\mathcal{L}_2(k^* + \Delta k)}{\mathcal{L}_2(k^* )} - 1 .
    \label{eq:delta_l2}
\end{equation}

By substituting $k^* = \frac{d}{e^{\epsilon} + 1}$, we have
\begin{multline*}
    \Delta \mathcal{L}_2 (\Delta k ) = \frac{(d-1) (e^{\epsilon} + 1)^4}{4d e^{\epsilon} - (e^{\epsilon } + 1)^2} \times
    \\ \begin{cases}
    \frac{1}{(2d + e^{\epsilon} + 1) (2d e^{\epsilon} - e^{\epsilon} - 1)} , \text{ if } \Delta k = \frac{1}{2} \\
    \frac{1}{(2d - e^{\epsilon} - 1) (2d e^{\epsilon} + e^{\epsilon} + 1)} , \text{ if } \Delta k = -\frac{1}{2} .
    \end{cases}
\end{multline*}
Since $2d < 2d e^{\epsilon}$, it follows that $(2d + e^{\epsilon} + 1) (2d e^{\epsilon} - e^{\epsilon} - 1) > (2d - e^{\epsilon} - 1) (2d e^{\epsilon} + e^{\epsilon} + 1)$. Therefore, $\Delta \mathcal{L}_2 (\Delta k) \le \frac{\mathcal{L}_2(k^* - \frac{1}{2})}{\mathcal{L}_2(k^*)} - 1$, which proves the $\mathcal{L}_2$ bound in Corollary \ref{cor:l1_l2_deviation}. Given Corollary \ref{cor:l1_to_l2}, the $\mathcal{L}_1$ bound is proven accordingly. $\qed$

\subsection{Proof of Corollary \ref{cor:l1_l2_k=1}}
\label{sec:proof_cor:l1_l2_k}
Substituting Eq. \eqref{eq:l2_k} with $k = 1$ proves the $\mathcal{L}_2$ term of this Corollary. Given Corollary \ref{cor:l1_to_l2}, the $\mathcal{L}_1$ bound is proven accordingly.

\section{Proof of Theorem \ref{thm:communication_cost_ub}}
\label{sec:proof_communication_cost_ub}
Recall that an optimal estimator requires an optimal symmetric configuration with an optimized support size (Definition \ref{def:osc}).

We now derive the upper bound for the number of rows in an optimal support matrix $S$, denoted by $m$. An optimal symmetric configuration requires $S^T P S = W$ to hold, where $P = \text{diag}(p_1, p_2, \dots, p_m)$ and $W$ is a $d \times d$ matrix such that for all $i \ne j$, $W_{i, j} = \frac{k(k-1)}{(d-1)[k(e^{\epsilon}-1) + d]}$, and for all $i$, $W_{i,i} = \frac{k}{k(e^{\epsilon}-1) + d}$. Without loss of generality and for the sake of concise representation, we normalize both sides of the equality by $\frac{d}{ \binom{d}{k} [k(e^{\epsilon}-1) + d]}$. After normalization, we have $\forall i \ne j: W_{i, j} = \frac{k(k-1)}{d(d-1)}$ and $\forall i: W_{i,i} = \frac{k}{d}$. Since $P$ is a diagonal matrix, its existing elements $p_o$ are scaled to $\frac{d p_o}{ \binom{d}{k} [k(e^{\epsilon}-1) + d]}$.

First, consider the base case where $S$ contains all possible rows with exactly $k$ ones (i.e., all $d$-choose-$k$ combinations). Denote this matrix as $S^{\star}$. When a row $S_{o,*}^{\star}$ is sampled uniformly from $S^{\star}$, the properties of combinations guarantee that $\forall i\ne j: \Pr(S_{o,i}^{\star} S_{o,j}^{\star} = 1) = \frac{k(k-1)}{d(d-1)}$ and $\forall i: \Pr(S_{o,i}^{\star} = 1) = \frac{k}{d}$. Consequently, $\forall i\ne j: \sum_o S_{o,i}^{\star} S_{o,j}^{\star}= \frac{k(k-1)}{d(d-1)} \binom{d}{k}$ and $\forall i : \sum_o S_{o,i}^{\star} = \frac{k}{d} \binom{d}{k}$. Let $P^{\star} = \frac{1}{ \binom{d}{k} } I$, where $I$ is the identity matrix. It follows that:
\begin{equation*}
    (S^{\star})^T P^{\star} S^{\star} = \frac{d}{ \binom{d}{k} [k(e^{\epsilon}-1) + d]} (S^{\star})^T S^{\star} = W .
\end{equation*}

We can reformulate $S^T P S = \sum_o p_o S_{o, *}^T S_{o, *}$, where $S_{o, *}$ is the $o$-th row of $S$. Since all diagonal elements $p_o$ are non-negative and $\sum_o p_o = 1$, $W$ is a convex combination of the set of matrices $\mathbf{C}^{\star} = \{(S_{o, *}^{\star})^T S_{o, *}^{\star} \mid o \in [ \binom{d}{k} ] \}$.

We now apply Carathéodory's theorem \cite{barvinok2025course} to prove that only $d(d-1) / 2 + 1$ rows of $S^{\star}$ are required to construct $W$. Carathéodory's theorem states that if a point $v$ is a convex combination of a set $V \subset \mathbb{R}^n$, then $v$ can be expressed as a convex combination of at most $n + 1$ points in $V$. Since $W$ is a convex combination of elements in $\mathbf{C}^{\star}$, it can be represented by a convex combination of a subset of $\mathbf{C}^{\star}$.

Next, we determine the required dimension $n$ by identifying the number of independent matrix elements needed to satisfy $W = \sum_o p_o S_{o, *}^T S_{o, *}$, where each $S_{o, *}^T S_{o, *}$ is an element of $\mathbf{C}^{\star}$.

First, because these matrices are symmetric (i.e., $W_{i, j} = W_{j, i}$ and $(S_{o, *}^T S_{o, *})_{i, j} = (S_{o, *}^T S_{o, *})_{j, i}$), the equality of $\text{SUT}(W) = \sum_o p_o \text{SUT}(S_{o, *}^T S_{o, *})$ automatically ensures the equality of the strict lower triangle. 

Second, we demonstrate that equality of the off-diagonal elements implies equality of the diagonal elements. Suppose all off-diagonal entries of the convex combination $\sum_o p_o S_{o, *}^T S_{o, *}$ match the target values in $W$, namely $\frac{k(k-1)}{d(d-1)}$. For any fixed index $i \in [d]$, we have:
\begin{multline*}
     \forall i \in [d]: \sum_{j \ne i} \sum_o p_o S_{o, i} S_{o, j} =  \sum_o p_o S_{o, i} \sum_{j \ne i}S_{o, j} \\ = \sum_o p_o S_{o, i}  (k-1) = (k-1) (p_o S_{o, *}^T S_{o, *})_{ii} 
\end{multline*}

\noindent where $\sum_{j \ne i} S_{o, j} = k-1$ because $S_{o, i}=1$ accounts for one of the $k$ ones in the row. Simultaneously, we have:
\begin{equation*}
      \sum_{j \ne i} \sum_o p_o S_{o, i} S_{o, j} =  (d-1) \sum_o p_o S_{o, i} S_{o, j} = (d-1)\frac{k(k-1)}{d(d-1)} ,
\end{equation*}

Equating these results yields $\forall i: (\sum_o p_o S_{o, *}^T S_{o, *})_{ii} = \frac{k}{d}$. Thus, matching the off-diagonal elements between $W$ and the convex combination automatically ensures the diagonal elements match.


By flattening the strictly upper triangle into a vector, we find its dimension is $d(d-1)/2$. Carathéodory's theorem ensures that at most $d(d-1)/2 + 1$ elements of $\mathbf{C}^{\star}$ are needed to reconstruct $W$. Therefore, we can construct a matrix $S$ for any given $k$ by stacking at most $d(d-1)/2 + 1$ rows. Since the communication cost is the logarithm of the number of possible responses (the rows of $S$), the cost is at most $\log_2 \left( \frac{d(d-1)}{2} + 1 \right)$.

\section{Proof of Optimized Count Mean Sketch}
\label{sec:proof_ocms}
A response $(h, z)$ of OCMS supports every element $x$ such that $h(x) = z$. Recall that $\mathcal{H}' = \{ h'(x) = (ax + b) \mod{d'} \mid a \in [1, d'-1], b \in [0, d'-1] \}$. Since $d'$ is a prime number, the properties of the Modular Multiplicative Inverse guarantee that each $h' \in \mathcal{H}'$ is equivalent to a permutation; specifically, $[h'(0), h'(1), \dots, h'(d-1)]$ is a permutation of $[0, 1, 2, \dots, d'-1]$. Furthermore, the Modular Multiplicative Inverse ensures that the pairs $(h_1'(x_1), h_1'(x_2))$ and $(h_2'(x_1), h_2'(x_2))$ are distinct if $h_1 \neq h_2$. Given that there are $d'(d'-1)$ distinct hash functions, we have:
\begin{lemma}
    Each hash function in $\mathcal{H}'$ maps an arbitrary pair $(x_1, x_2)$ to all possible pairs $(y_1, y_2)$ exactly once.
    \label{lem:uniform_pairwise_distinctness}
\end{lemma}

The operation $h'(x) \mod{B}$ then maps these permuted elements into $B$ buckets. Since $d'$ is not divisible by $B$, some responses support $\lceil \frac{d'}{B} \rceil$ elements while others support $\lfloor \frac{d'}{B} \rfloor$ elements.

\subsection{Proof of Theorem \ref{thm:cms_equivalance}}
\label{sec:proof_thm:cms_equivalance}
The support matrix $S$ of OCMS consists of $d'(d'-1)B$ rows. Given that the hash function $h$ is sampled uniformly and the randomized response has a constant base probability, the base probability of $S$ is also constant. Although Lemma \ref{lem:sp_and_p} assumes a constant support size, it generalizes to any symmetric configuration where $k = \frac{\sum_o p_o k_o}{\sum_o p_o}$. For OCMS, since $k = d'/B$, the base probability is $p_o = \frac{d'}{\frac{d'}{B} (e^{\epsilon} - 1) + d'} \cdot \frac{1}{Bd'(d'-1)}$.

We construct the support matrix of the first configuration, $S^{(1)}$, by stacking all rows of $S$ where the response $(h, z)$ satisfies $z \le r = d' \mod B$. Since the buckets corresponding to $z \le r$ contain $\lceil \frac{d'}{B} \rceil$ elements, all rows of $S^{(1)}$ have a constant support size of $\lceil d'/B \rceil $. Defining the base probability of $S^{(1)}$ as a constant, Lemma \ref{lem:sp_and_p} yields $p_1 = \frac{d'}{(\frac{d' - r}{B} + 1) (e^{\epsilon} - 1) + d'} \cdot \frac{1}{r d' (d' - 1)}$.

Similarly, the support matrix of the second configuration, $S^{(2)}$, is constructed from rows where $z > r$, resulting in a constant support size of $\lfloor \frac{d'}{B} \rfloor$. Its base probability is computed as $p_2 = \frac{d'}{\frac{d' - r}{B} (e^{\epsilon} - 1) + d'} \cdot \frac{1}{(B - r) d' (d' - 1)}$. It can be verified that $p_{\alpha} p_1 = p_0$ and $(1 - p_{\alpha}) p_2 = p_0$, demonstrating that the sampling of these two configurations generates the same support matrix $S$.

These split configurations possess constant support sizes and extremal configurations. We now prove that their support matrix and base probability satisfy the requirements of a symmetric configuration. Taking $S^{(1)}$ as an example, this requires $p_1 \sum_o S^{(1)}_{o, i} S^{(1)}_{o, j}$ to be constant for all $i \neq j$. For an arbitrary pair $(i, j)$ where $i \neq j$, the term $\sum_o S^{(1)}_{o, i} S^{(1)}_{o, j}$ represents the number of hash functions mapping $(i, j)$ into the same bucket of size $\lceil \frac{d'}{B} \rceil$. Elements within such a bucket can form $(\lceil \frac{d'}{B} \rceil + 1)(\lceil \frac{d'}{B} \rceil)$ distinct pairs, and there are $r$ such buckets. Mapping to the same bucket is equivalent to mapping to these distinct pairs. By Lemma \ref{lem:uniform_pairwise_distinctness}, there are exactly $r (\lceil \frac{d'}{B} \rceil + 1)(\lceil \frac{d'}{B} \rceil)$ hash functions mapping $(i, j)$ to these pairs. This holds for every $(i, j)$, ensuring the sum is constant.

The same logic applies to $S^{(2)}$, confirming both configurations are symmetric. Since the sampling between these two configurations generates an equivalent support matrix to OCMS, Theorem \ref{thm:cms_equivalance} is proved. $\qed$

Furthermore, extending this counting method to $B$ buckets and ignoring the remainder $r$, one can derive $P(h(x_1) = h(x_2)) \approx \frac{d'-B}{B^2(d'-1)}$, verifying the reduction in collision probability discussed in Section \ref{sec:ocms}.

Finally, if a process generates variable $V_1$ with probability $p$ and variable $V_2$ with probability $1-p$, and $E[V_1] = E[V_2]$, the variance of the process is $p \text{Var}(V_1) + (1-p) \text{Var}(V_2)$. This proves Corollary \ref{cor:cms_var}. $\qed$

\subsection{Proof of Corollary \ref{cor:cms_deviation}}

We begin by analyzing the impact of extending the dictionary size from $d$ to $d'$. When the dictionary size is increased, the $\mathcal{L}_2$ loss does not need to account for the extended values, as they are known to be identically zero. Thus, the $\mathcal{L}_2$ loss with an extended dictionary is given by:

\begin{equation*}
    \mathcal{L}_2'(\hat{f}) = \sum_{x \in [d]} \text{Var}(\hat{f}(x) |  d') ,
\end{equation*}

\noindent where $\text{Var}(\hat{f}(x) \mid d')$ denotes the variance calculated with $d'$ in place of $d$. Assuming an optimal estimator (with the $k$ term addressed in the subsequent step), we set $k = \frac{d'}{e^{\epsilon} + 1}$ to optimize $\mathcal{L}_1$ and $\mathcal{L}_2$. Given Eq. \eqref{eq:l2_k}, we obtain:
\begin{equation*}
    \frac{\mathcal{L}_2'(\hat{f})}{\mathcal{L}_2(\hat{f})} - 1 = \frac{(d' - d) (e^{\epsilon} + 1)^2 (2d d' - d - d')}{d'^2 (d - 1) (4de^{\epsilon} - (e^{\epsilon} + 1)^2)} = \beta
\end{equation*}

Next, we evaluate the deviation of $k$ from the optimal value $k^*$ for a given dictionary size $d'$. If $B = \frac{d'}{k^*}$ is an integer, OCMS is equivalent to sampling between the configurations with support size of $\lceil k^* \rceil$ and $\lfloor k^*\rfloor$. If $B = \frac{d'}{k^*}$ is not an integer, then $B$ is either $\lceil \frac{d'}{k^*} \rceil$ or $\lfloor \frac{d'}{k^*} \rfloor$, which correspond to sampling between configurations with support size of either ($k^* - 1$ and $k^*$) or ($k^* + 1$ and $k^*$). In all such scenarios, the support size of the sampled configurations deviates from $k^*$ by at most one. Analogous to Eq. \eqref{eq:delta_l2}, we define the relative change in variance as:

\begin{equation*}
    \Delta \mathcal{L}_2' (\Delta k) = \frac{ \mathcal{L}_2' (\frac{d'}{e^{\epsilon} +1}+ \Delta k) }{\mathcal{L}_2'(\frac{d'}{e^{\epsilon} +1})} - 1 .
\end{equation*}

It can be verified that $\Delta \mathcal{L}_2' (\Delta k)$ is a decreasing function of $d'$; therefore, it is upper-bounded by the case where $d' = d$. Consequently, we have:
\begin{multline*}
    \Delta \mathcal{L}_2 (\Delta k ) = \frac{(d-1) (e^{\epsilon} + 1)^4}{4d e^{\epsilon} - (e^{\epsilon } + 1)^2} \times
    \\ \begin{cases}
    \frac{1}{(d + e^{\epsilon} + 1) (d e^{\epsilon} - e^{\epsilon} - 1)} , \text{ if } \Delta k = 1 \\
    \frac{1}{(d - e^{\epsilon} - 1) (d e^{\epsilon} + e^{\epsilon} + 1)} , \text{ if } \Delta k = -1 .
    \end{cases}
\end{multline*}

Since $d < d e^{\epsilon}$, it follows that $(d + e^{\epsilon} + 1) (d e^{\epsilon} - e^{\epsilon} - 1) > (d - e^{\epsilon} - 1) (d e^{\epsilon} + e^{\epsilon} + 1)$. Therefore, $\Delta \mathcal{L}_2 (\Delta k ) \le \frac{\mathcal{L}_2(k - 1)}{\mathcal{L}_2(k)} - 1$, where $\frac{\mathcal{L}_2(k - 1)}{\mathcal{L}_2(k)} - 1$ corresponds to $\alpha$ in Corollary \ref{cor:cms_deviation}. The deviation of $\mathcal{L}_1$ follows directly from Corollary \ref{cor:l1_to_l2}. $\qed$

\end{document}